\documentclass{aa}

\usepackage{graphicx}
\usepackage{txfonts}
\usepackage{natbib}

\bibpunct{(}{)}{;}{a}{}{,}

\newcommand\bb[1] {   \mbox{\boldmath{$#1$}}  }

\begin{document}

\title{Global MHD simulations of stratified and turbulent
  protoplanetary discs. II. Dust settling}
\author{S\'ebastien Fromang \inst{1,2} and Richard P. Nelson \inst{3}}

\offprints{S.Fromang}

\institute{CEA, Irfu, SAp, Centre de Saclay, F-91191 Gif-sur-Yvette, France \and UMR AIM, CEA-CNRS-Univ. Paris VII, Centre de Saclay, F-91191
Gif-sur-Yvette, France. \and Astronomy Unit, Queen Mary, University of London, 
Mile End Road, London E1 4NS, UK \\ \email{sebastien.fromang@cea.fr}}

\date{Accepted; Received; in original form;}

\label{firstpage}

\abstract
{}
{The aim of this paper is to study the vertical profile
 of small dust particles in protoplanetary discs in which
 angular momentum transport is due to MHD turbulence driven by the
 magnetorotational instability. We consider particle sizes that range
 from approximately $1$ micron up to a few millimeters.} 
{We use a grid--based MHD code to perform global two-fluid
 simulations of turbulent protoplanetary discs which contain
 dust grains of various sizes.}
{In quasi--steady state, the gravitational settling of dust particles is
 balanced by turbulent diffusion. Simple and standard models of this
 process fail to describe accurately the vertical profile of the
 dust density. The disagreement is larger for small dust particles
 (of a few microns in size), especially in the disc upper layers
 ($Z>3H$, where $H$ is the scale-height). 
 Here there can be orders of magnitude in the disagreement 
 between the simple model predictions and the simulation results. 
 This is because MHD turbulence is not homogeneous in accretion 
 discs, since velocity fluctuations increase significantly in the disc 
 upper layer where a strongly magnetized corona develops.
 We provide an alternative model that gives a better fit to the
 simulations. In this model, dust particles are diffused away from 
 the midplane by MHD turbulence, but the diffusion coefficient 
 varies vertically and is everywhere proportional to the square of the
 local turbulent vertical velocity fluctuations.}
{The spatial distribution of dust particles can be used to trace
  the properties of MHD turbulence in protoplanetary discs, such as
  the amplitude of the velocity fluctuations. In the
  future, detailed and direct comparison between numerical simulations
  and observations should prove a useful tool for constraining
  the properties of turbulence in protoplanetary discs.}

\authorrunning{S.Fromang \&  R.Nelson}
\titlerunning{Dust settling in turbulent protoplanetary discs}
\maketitle

\section{Introduction} 
\label{intro}

Thanks to the Spitzer Space Telescope, the last few years have seen a
dramatic improvement in our knowledge of dust emission features  
arising at mid-infrared wavelengths from protoplanetary discs
surrounding brown dwarfs, T Tauri stars and Herbig Ae/Be stars. The
properties (size, composition) of these grains can be studied in
great detail by analyzing the shape and strength of these emission
features. The main result is that dust particles have been
significantly processed compared to their interstellar medium cousins:
grains are bigger (up to a few microns) and, for reasons not yet fully
understood, crystallinity appears to be common among all observed
spectral types \citep{apaietal05,kessleretal06,furlanetal06}. Because
protoplanetary discs are optically thick, this mid-infrared 
emission mostly arises from the upper layers of the disc
\citep{vanboekeletal03,dullemond&dominik04,dullemond&dominik08},
in the so--called ``superheated'' layer of \citet{chiang&goldreich97},
and is attributed to the disc inner radii. In the case of T Tauri
stars of solar-type, for example, the dust emission zone lies within a
few tenths of an astronomical unit (AU) of the central star
\citep{kessleretal07}.  

The presence of solid particles high above the disc midplane is the
signature of the turbulent nature of the flow in
protoplanetary discs. Turbulent velocity fluctuations lift up dust
particles that would otherwise settle toward the
disc midplane \citep{dubrulleetal95,dullemond&dominik04} because of
the vertical component of the central star's gravitational
potential. The vertical profile of the dust density
is thus determined by the balance between gravitational
settling toward the equatorial plane and upward lift due to
turbulence. Its typical scale-height is a function
of grain size. Small particles are well coupled to the gas, and
can be transported to higher altitudes (or smaller gas 
densities) above the equatorial plane than larger particles.
Although the issue is still under discussion
\citep{dullemond&dominik08}, this differential settling
process can be inferred from a detailed analysis of the
observations. This is the case for observations carried out by 
Spitzer alone \citep{furlanetal05,siciliaaguilaretal07}, by
combining Spitzer observations with observations at other wavelengths
\citep{pinteetal08} or even using completely different observational
strategies \citep{pinteetal07,rettigetal06}. Dust gravitational
settling is also known to affect the spectral energy 
distribution of protoplanetary discs, as shown for example by
\citet{dullemond&dominik04} or \citet{dalessioetal06}. 

In general, such observational diagnostics of gravitational settling
are inferred 
using simple parametric prescriptions of the effect of turbulence or of its
consequence. \citet{pinteetal07,pinteetal08} assume that the vertical
profile of the dust density is a Gaussian (as is the case for the gas
in the isothermal limit), leaving open the possibility that the dust
disc scale-height depends on the size of the particles. The point of
their analysis is to demonstrate that this is indeed the
case in GG Tau and IM Lupi. \citet{furlanetal05} use the models developed by
\citet{dalessioetal06} and assume a constant dust--to--gas ratio for
two populations of solids, one consisting of small and well mixed
particles, the other composed of large particles. Using Spitzer
spectra, they demonstrate that the latter is depleted in the disc upper
layers for the vast majority of a sample of 25 stars in
Taurus. \citet{dullemond&dominik04} use a more 
physical approach in which
turbulence is modeled as a diffusive process with a spatially constant
diffusion coefficient 
\citep[see also][]{dubrulleetal95,schapler&henning04}.
Using this formulation, one can derive
analytic expressions for the dust density vertical
profile. This is also the approach followed by \citet{rettigetal06}
who used the strong settling (thin dust disc) limit of these formulae.

All of these approaches are very useful since they provide analytical
formulae or a small number of model parameters to fit. They can
thus be incorporated into radiative transfer tools, which
can then generate a large
grid of models to be used for interpreting the observations. All of them 
establish, at least qualitatively if not quantitatively, the basic
result that differential gravitational settling occurs in protoplanetary
discs. However, none of these derivations follows directly
from first principles, but instead rely at best on a set of
unchecked assumptions concerning the properties of the underlying
turbulence. The origin of the latter, however, is thought to
be magnetohydrodynamic in nature and driven by the magnetorotational
instability \citep[MRI,][]{balbus&hawley91,balbus&hawley98}. It is now
possible to perform global numerical simulations of turbulent
protoplanetary discs, 
including solid particles, and to use these simulations as a test of
the models described above. This is the purpose of the present
paper. Despite the recent increase in computational resources, it is
important to keep in mind that such simulations are still extremely
challenging and need to use as simplified a set-up as possible. This
is not without consequences for their realism. Important 
issues related to the MRI, such as small scale dissipation
\citep{fromangetal07, lesur&longaretti07} and the possible presence of
a dead zone \citep{gammie96} still have to be ignored. We will return
to these issues in the last section of the paper when discussing
the limits of our work. 

Of course, the effect of MRI--induced MHD turbulence on dust dynamics
and dust vertical settling itself has already been studied in numerical
simulations, but this is the first time that both effects are
incorporated in a single global simulation that takes vertical
stratification into account. \citet{barriereetal05} performed
global simulations of dust settling but neglected the effect of MHD
turbulence. This prevents the system from reaching a quasi--steady
state. Other global simulations, taking MRI--driven turbulence into
account and including dust particles, neglected gaseous vertical
stratification \citep{fromang&nelson05} and instead focussed on the radial
migration of larger bodies. \citet{lyraetal08} also neglected the
vertical component of gravity acting on the gas, but included its
effects on the dust particles. While this approach produces settling
of the dust particles toward the midplane, its neglects the spatial
inhomogeneity of the turbulence induced by vertical stratification
of the gas. We shall see in the course of this paper that the latter
is important when considering dust settling of small particles. The
other published numerical simulations studying the effect of
turbulence on dust dynamics were local simulations that
use the shearing box model \citep{goldreich&lyndenbell65}. A large
number of them neglected density stratification in the vertical
direction \citep{johansenetal05,carballidoetal05,johansenetal06} and
thus only studied dust diffusion in an homogeneous
environment. Others \citep{fromang&pap06,carballidoetal06} included
vertical density stratification but considered particles 
larger than one millimeter. Here, we want to
concentrate on smaller particles that produce an observational
signature at mid--infrared wavelengths. We note that such a
simulation could in principle be done in the framework of the shearing
box model. Indeed, this was done recently by
  \citet{balsaraetal08}, although the vertical extent of their
  shearing box is
  smaller than the simulations we present in this paper and thus less
  appropriate to study the dust distribution in the disc corona
  (i.e. above three scale-heights). However, since this work is intended
to mark the beginning 
of an effort to compare observations and numerical simulations
directly, it makes more sense to compute global models as
these will be more readily comparable with the observations as they
become more realistic. Our strategy in this paper will be simple. We
will use exactly the set-up presented by \citet{fromang&nelson06} for
global simulations of turbulent and stratified protoplanetary
discs. Dust particles of various sizes will be added to the
disc and their subsequent evolution will be analyzed
and compared with simple models of dust stratification in
protoplanetary discs.

The plan of the paper is as follows. In Sect.~\ref{definitions}, we
introduce the model we use for the disc as well as 
convenient dimensionless
parameters that appear in the problem. The relationship between 
these quantities and physical parameters in real systems will be
outlined. In Sect.~\ref{dust_settling_sec}, we describe more
quantitatively the different ways to model the quasi--steady state
dust distribution resulting from the balance between dust settling and
turbulent diffusion. These models are then compared with the results of
our numerical simulations in Sect.~\ref{simu_sec}. Finally,
in Sect.~\ref{conclusion_section} we discuss the implications and
limitations of our work, and point the way toward future
improvements.

\section{Definitions}
\label{definitions}

In this section, we describe the general properties of the disc model
and the dust parameters we used. We also introduce the 
mathematical notation that will be required in the following sections.

\subsection{Coordinate systems}
\label{coord_syst}

In this paper, we will use both cylindrical and
  spherical coordinate systems. The former will be denoted by
  ${(R,\phi,Z)}$, and will be used mostly in the present 
  Sect.~\ref{definitions} and in Sect.~\ref{dust_settling_sec}. Spherical
  coordinates will be used when we describe the numerical simulations
  and will use the notation $(r,\theta,\phi)$.

\subsection{Disc model}
\label{disc_model_sec}

We consider a disc extending 
in radius between an inner radius $R_{d,in}$ and an
outer radius $R_{d,out}$. For simplicity and computational
reasons we define the initial disc structure using straight-forward
analytic functions. The equation of state is locally isothermal:
the sound speed, $c_s$, only depends on $R$ and is
constant in time. Both $c_s$ and the disc midplane gas density,
$\rho_m$, obey power laws
\begin{eqnarray}
  c_s^2(R)&=&c_0^2 \left( \frac{R}{R_0} \right)^{-1} \,
  , \label{cs_def} \\
\rho_{mid}(R)&=&\rho_0 \left( \frac{R}{R_0} \right)^{-3/2} \, , \label{rho_def}
\end{eqnarray}
where $c_0$ and $\rho_0$ are the sound speed and the
midplane gas density at a radius $R_0$, respectively. 
The disc is initially axisymmetric and in radial and 
vertical hydrostatic equilibrium. The
spatial distribution of density $\rho(R,Z)$ and angular velocity
$\Omega(R,Z)$ can thus be approximated by
\begin{eqnarray}
\rho(R,Z) &=&\rho_{mid}(R)  e^{-Z^2/2H^2}=\rho_0 \left( \frac{R}{R_0}
\right)^{-3/2} e^{-Z^2/2H^2} \, , \\
\Omega(R,Z) &=&\sqrt{\frac{GM}{R^3}}=\Omega_0 \left( \frac{R}{R_0}
\right)^{-3/2} \, , 
\end{eqnarray}
in which $\Omega_0=\sqrt{GM/R_0^3}$. 
The disc scale-height, $H$, is given by
\begin{equation}
H=\frac{c_s}{\Omega}=H_0 \left( \frac{R}{R_0} \right) \, ,
\end{equation}
where $H_0=c_0/\Omega_0$ is the disc scale-height at $R_0$.

\subsection{Integrated quantities}

The density distribution can be used to work out the surface density of the
disc:
\begin{equation}
\Sigma(R) = \int_{-\infty}^{+\infty} \rho dZ = \sqrt{2\pi} \rho_0 \frac{c_0}{\Omega_0} \left( \frac{R_0}{R} \right)^{1/2} = \Sigma_0 \left( \frac{R_0}{R} \right)^{1/2} \, ,
\label{sigma_eq}
\end{equation}
where the second relation serves as a definition for $\Sigma_0$. The total disc 
mass $M_d$ follows by radially integrating the surface density between
$R_{d,in}$ and $R_{d,out}$:
\begin{equation}
M_d=\frac{4\pi}{3}\Sigma_0 R_0^2 \left( \frac{R_{d,out}}{R_0} \right)^{3/2} \, ,
\end{equation}
where we have assumed $R_{d,in} \ll R_{d,out}$. This relation can be used
to express the disc surface density at $R_0$ as a function of the disc
parameters and $R_0$:
\begin{equation}
\Sigma_0=\frac{3M_d}{4\pi R_0^2}\left( \frac{R_0}{R_{d,out}}
\right)^{3/2} \, .
\label{sigmamass_eq}
\end{equation}

\subsection{Dust size}

In this paper, we shall study the effect of MHD turbulence on the
dust. Gas affects solid body dynamics through the drag force it exerts
on the dust particles. In the Epstein regime we are interested in,
this force $\bb{F_d}$ takes the simple form
\begin{equation}
\bb{F_d}=-\frac{\bb{v}-\bb{v_d}}{\tau_s} \, ,
\end{equation}
where $\bb{v}$ and $\bb{v_d}$ are the gas and dust
velocities, respectively. $\tau_s$ is the dust stopping 
time. This is the typical time it
takes for dust particles initially at rest to reach the local gas
velocity. It depends on the dust particle mass density $\rho_s$ and
its size $a$ through
\begin{equation}
\tau_s=\frac{\rho_s a}{\rho c_s} \, .
\end{equation}
A relevant dimensionless parameter in the problem is the quantity 
$\Omega \tau_s$. It compares the stopping time to the dynamical
time. When $\Omega \tau_s \ll 1$, the stopping time is much smaller
than the orbital period $T_{orb}$ and the dust essentially follows the gas. When
$\Omega \tau_s \sim 1$ or larger, $\tau_s$ becomes comparable to $T_{orb}$ 
and dust and gas start to decouple. As pointed out by
\citet{dullemond&dominik04}, this occurs at all radii for a given
particle size $a$ provided $Z$ is large enough. Using the disc
parameters introduced above, $\Omega \tau_s$ can be expressed as a
function of position according to
\begin{equation}
\Omega \tau_s=\Omega_0 \frac{\rho_s a}{\rho_0 c_0} \left(
\frac{R}{R_0} \right)^{1/2} e^{Z^2/2H^2} =(\Omega \tau_s)_0 \left(
\frac{R}{R_0} \right)^{1/2} e^{Z^2/2H^2} \, ,
\end{equation}
where the parameter $(\Omega \tau_s)_0$ is the value of $\Omega \tau_s$
at $R=R_0$ in the disc midplane. It can be expressed in terms of the
disc surface density at $R_0$ using Eq.~(\ref{sigma_eq}):
\begin{equation}
(\Omega \tau_s)_0=\sqrt{2 \pi} \frac{\rho_s a}{\Sigma_0}.
\end{equation}
Using this expression along with Eq.~(\ref{sigmamass_eq}), it is
possible to express the dust size in term of $(\Omega \tau_s)_0$, the
disc parameters (mass and radius) and $R_0$:
\begin{equation}
a=\frac{3M_d}{2\sqrt{2\pi}\rho_sR_0^2}\left(
\frac{R_{d,out}}{R_0} \right)^{-3/2} (\Omega \tau_s)_0.
\label{dustsize_eq}
\end{equation}

\subsection{Converting to physical units}

Eq.~(\ref{sigmamass_eq}) and (\ref{dustsize_eq}) can be used to
convert the dimensionless quantities describing the problem into
physical quantities. When doing so, the numerical simulations we 
will describe in
Sect.~\ref{simu_sec} should be thought of as covering a small
fraction of the total disc, going from $R_{in}=R_0 \geq
R_{d,in}$ to an outer radius $R_{out} \leq R_{d,out}$.

The dimensionless parameters describing the disc and dust particles,
and the results of the numerical simulations presented below, can
be rescaled to any physical system upon specifying the disc mass,
the outer radius of the disc, 
and the value of $R_0$ (in astronomical units).
For example, the disc surface density 
$\Sigma_0$ can be written as
\begin{equation}
\Sigma_0=4 \left( \frac{M_d}{0.01 M_\odot} \right) \left( \frac{R_{d,out}}{300
  AU}\right)^{-3/2} \left( \frac{R_0}{1 AU} \right)^{-1/2} {\rm g} 
  {\rm cm}^{-2}.
\label{sigma_phys}
\end{equation}
Likewise, taking $\rho_s=1$ gcm$^{-3}$ and using
Eq.~(\ref{dustsize_eq}), we obtain an expression for the dust size:
\begin{equation}
a= 163 \left( \frac{(\Omega \tau_s)_0}{0.01} \right) \left(
\frac{M_d}{0.01 M_\odot} \right) \left( \frac{R_{d,out}}{300 AU}
\right)^{-3/2} \left( \frac{R_0}{1 AU} \right)^{-1/2} \mu m \, .
\label{dustsize_phys}
\end{equation}

In Sect.~\ref{simu_sec}, we will describe the results of three
simulations. They are characterized by $(\Omega
\tau_s)_0=10^{-2}$,  $10^{-3}$ and $10^{-4}$. In a disc of mass $0.01
M_\odot$ and $300$ AU in size, they would correspond to
dust particles of size $163$, $16$ and $1.6$ $\mu$m, respectively,
if we take $R_0=1$ AU.
(This would mean that we consider the simulation to cover the
radial extent $1$ to $8$ AU.) For the same disc mass and size, if we
now take $R_0=0.1$ AU (i.e. we consider the simulation to cover radii
ranging from $0.1$ to $0.8$ AU), the three sizes are $500$,
$50$ and $5$ $\mu$m. Note, however, that these numbers are only
illustrative. In general, for a
given value of $R_0$, the size of the dust particles decreases when
the disc mass is decreased or its outer radius is increased.

\section{Dust settling in turbulent discs}
\label{dust_settling_sec}

As pointed out in the introduction, a steady state is reached in a
turbulent protoplanetary disc in which turbulent fluctuations oppose
and balance against dust settling. In this section, we describe three
approaches that can be used to model the vertical profile of the
dust density.

\subsection{A Gaussian profile}
\label{gauss_prof_sec}

The simplest approach, and one that is commonly used when attempting
to interpret observations \citep{pinteetal08}, 
is to assume that the dust density
follows a Gaussian distribution, as the gas does, but with a
different vertical scale-height $H_d$:
\begin{equation}
\rho_d=\rho_{d,mid}  e^{-Z^2/2H_d^2}\, ,
\end{equation}
where $\rho_{d,mid}$ is the dust midplane density. In this approach,
$H_d$ is different for each dust particle size $a$. For example,
while trying to model the dust properties in the protoplanetary disc
orbiting the M dwarf IM Lupi, \citet{pinteetal08} found $H_d
\propto a^{-0.05}$. One purpose of this paper is to establish
whether such a description is supported by numerical simulations of
turbulent protoplanetary discs.

\subsection{Turbulence as a diffusive process}

A more physical approach is to describe the transport of the
dust particles by the turbulent fluctuations as a diffusion
process. This has been used commonly in the literature
\citep{dullemond&dominik04}. In this approach, the vertical evolution
of the dust density can be described by the following 
partial differential
equation \citep{schapler&henning04,dubrulleetal95}: 
\begin{equation}
\frac{\partial \rho_d}{\partial t}-\frac{\partial}{\partial z}(z
\Omega^2 \tau_s \rho_d)= \frac{\partial}{\partial z} \left[ D \rho 
  \frac{\partial}{\partial z} \left( \frac{\rho_d}{\rho} \right)
  \right] \, ,
\label{diff_advec_eq}
\end{equation}
where $\rho_d$ is the dust particle density and $D$ is 
a diffusion coefficient that quantifies the turbulent diffusivity.
This equation models the balance between vertical 
settling and turbulent diffusion. When looking for a steady state
vertical profile for the density, the time derivative vanishes. Upon
integrating once and rearranging terms, Eq.~(\ref{diff_advec_eq})
gives
\begin{equation}
\frac{\partial}{\partial z} \left( \ln
\frac{\rho_d}{\rho}\right)=-\frac{\Omega^2\tau_s}{D} z \, .
\label{diff_advec_eq_II}
\end{equation}
The vertical integration of the last equation requires the knowledge
of the diffusion coefficient as a function of $z$. The simplest
solution is to assume that it is constant. This is the approach
described in the following subsection, while in
Sect.~\ref{varying_D} we outline a possible alternative.

\subsubsection{A constant diffusion coefficient}
\label{const_D}

When the dust diffusion coefficient $D$ is constant,
Eq.~(\ref{diff_advec_eq_II}) can be integrated to give
\begin{equation}
\rho_d=\rho_{d,mid}
\exp \left[-\frac{(\Omega\tau_s)_{mid}}{\tilde{D}} \left(
  exp\left(\frac{Z^2}{2H^2}\right)-1\right) -\frac{Z^2}{2H^2} \right]
\label{dust_prof_eq}
\end{equation}
where  $\tilde{D}$ is a dimensionless diffusion coefficient defined as
$D=\tilde{D}c_s H$. The quantities $\rho_{d,mid}$ and
$(\Omega\tau_s)_{mid}$ only 
depend on $R$ and are to be evaluated in the disc midplane. Note
that, while deriving the last equation, we have assumed that the
vertical distribution of gas remains Gaussian at all times, in
agreement with local shearing box numerical simulations of the MRI
\citep{miller&stone00}. 

A common practice in this context is to express $\tilde{D}$
in units of the standard $\alpha$ parameter introduced by
\citet{shakura&sunyaev73}. $\tilde{D}$ is then written as follows
\citep{dullemond&dominik04,schapler&henning04}:
\begin{equation}
\tilde{D}=\frac{\alpha}{\textrm{Sc}}
\label{const_D_eq}
\end{equation}
where $\textrm{Sc}$ is the Schmidt number. In zero net flux MHD turbulence, the
Schmidt number has been measured to be of order unity in local
simulations of unstratified \citep{johansen&klahr05,johansenetal06} or
stratified discs \citep{fromang&pap06,ilgner&nelson08}. 
In the present paper, we will
tune the Schmidt number in order to obtain the best agreement with the
numerical simulations.

\subsubsection{A vertically varying diffusion coefficient}
\label{varying_D}

Dust particles are diffused away from the disc midplane by the
turbulent velocity fluctuations.
Thus, the dust diffusion coefficient is intimately linked
to the turbulence properties and particularly to the gas velocity
fluctuations. It would then seem natural for $D$ to be constant in
space if the turbulence was homogeneous. However, because of the
vertical stratification, MHD turbulence is not homogeneous in
protoplanetary discs and it is very likely that $D$ varies with $Z$ at
a given radius (even in the absence of a dead zone). 
Using local vertically stratified simulations of the
MRI, \citet{fromang&pap06} showed that the following
simple expression gives a fairly good estimate to the numerically
derived diffusion coefficient:
\begin{equation}
D=\delta v_z^2 \tau_{corr} \, .
\label{D_vel}
\end{equation}
In this equation, $\delta v_z$ stands for the turbulent velocity
fluctuations and $\tau_{corr}$ is the correlation time of these
fluctuations. Numerical estimates of both terms thus provide a path
to calculating the value of $D$. Their vertical variations will be
investigated in Sect.~\ref{simu_sec}.

Of course, the drawback of this approach is that it becomes impossible
to explicitly integrated Eq.~(\ref{diff_advec_eq_II}). We
cannot provide an analytical expression for the vertical profile of
the dust density and shall rely on a numerical integration once the
vertical profile for $D$ is extracted from the numerical simulations.

\section{Numerical simulations}
\label{simu_sec}

\subsection{Set-up}

\begin{figure*}
\begin{center}
\includegraphics[scale=0.4]{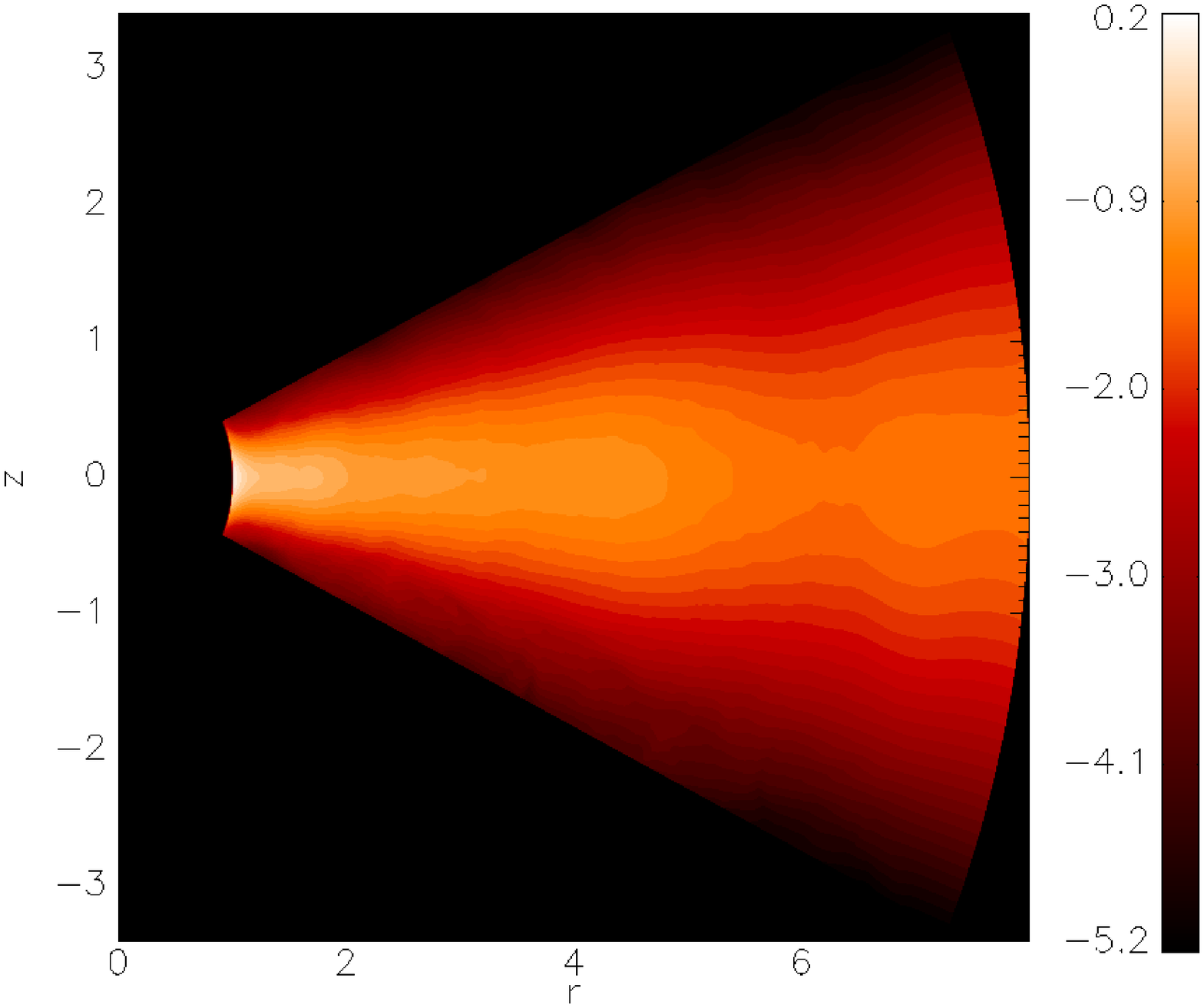}
\includegraphics[scale=0.4]{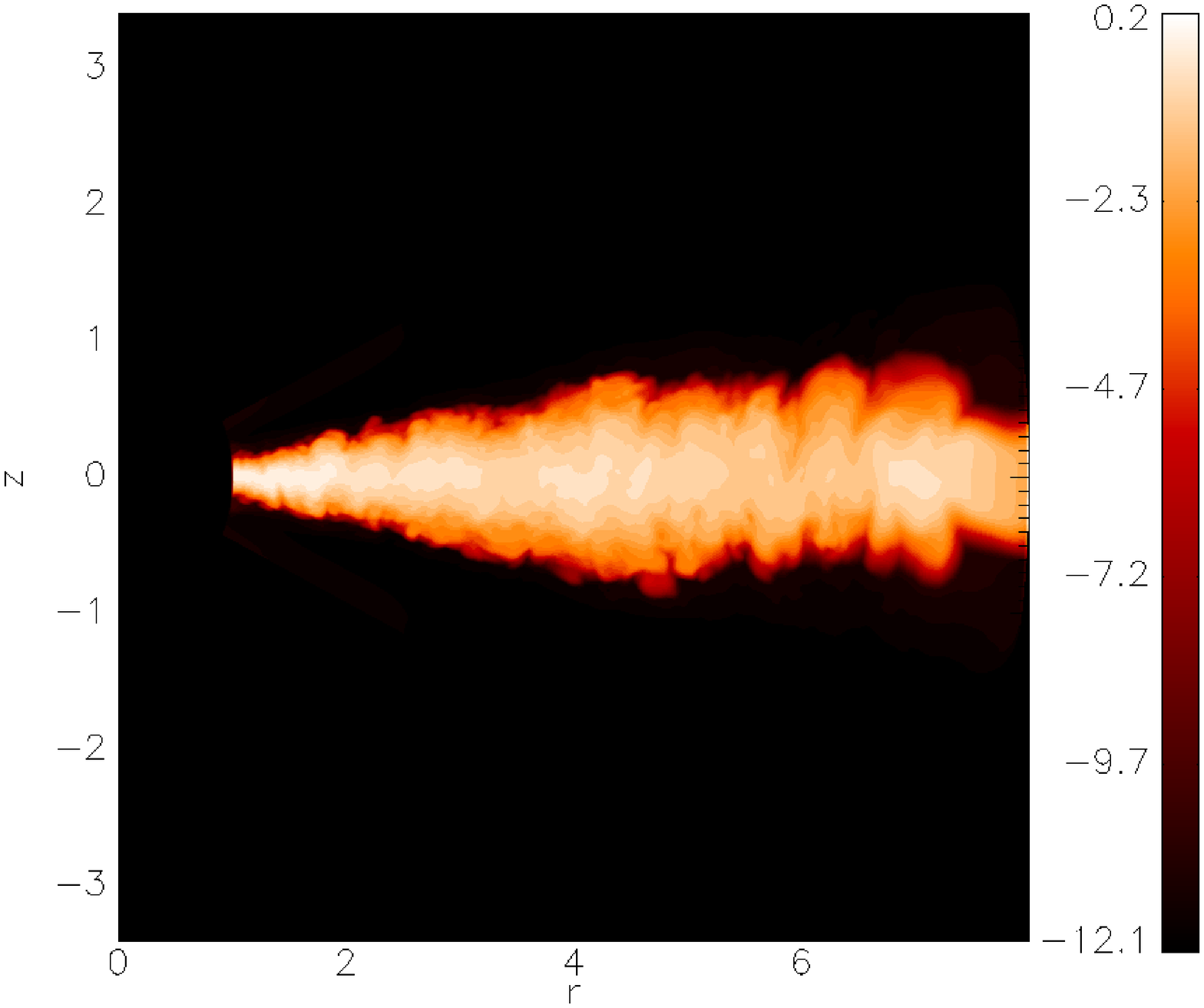}
\caption{Snapshots of the gas density ({\it left panel}) and the dust 
density ({\it right panel}) at $t=600$ for the case 
$(\Omega \tau)_0=0.01$.}
\label{snap_0.01}
\end{center}
\end{figure*}

The simulations presented in this paper are run using the code GLOBAL
 \citep{hawley&stone95}, which solves the ideal MHD equations
   using a spherical coordinate
   system as defined in Sect.~\ref{coord_syst}. The set-up we used
 is exactly that of model S5 
in \citet{fromang&nelson06}. Here, we describe it only briefly. 

At the start of the simulation, the disc model presented in
Sect.~\ref{disc_model_sec} is initialized on the grid. The units are
such that $GM=1$, $c_0=0.1$ and $\rho_0=1$ (i.e. $H/R=0.1$ throughout
the disc). The computational domain covers the range $R_{in}=R_0=1$ to
$R_{out}=8$ in radius and the interval $[0,\pi/4]$ in $\phi$. 
In the $\theta$-direction, the grid extends to 
$4.3$ scale-heights on both sides of
the disc equatorial plane. The resolution for each simulation is
$(N_r,N_{\phi},N_{\theta})=(364,124,213)$. Following
\citet{fromang&nelson06}, a weak toroidal magnetic field is added to
the disc, and small random velocity perturbations are also 
imposed. Time is measured in units of the
orbital period at the inner edge of the computational domain in the
following sections.

Because of the MRI, the presence of a weak magnetic field, together
with the velocity perturbations, 
begins to drive MHD turbulence and angular momentum
transport through the disc within a few orbits of the simulations
starting. To reach a meaningful quasi
steady state, however, the model is first evolved for $430$ orbits
without dust particles. At that stage, the gas density is reset to its
initial distribution and dust 
particles are introduced such that the dust-to-gas ratio is uniform
through the computational domain (note that we neglect the back
reaction of the solids onto the gas, so that the value of this ratio 
has no physical consequences). We ran three simulations for 
three different particle
sizes. The three values of $(\Omega \tau_s)_0$ associated with these
sizes are $10^{-2}$, $10^{-3}$ and $10^{-4}$. All the simulations are
further integrated for about $200$ orbits until the dust distribution
itself reaches a steady state in which gravitational settling is
balanced by turbulent diffusion. Examples of the disc structure 
at the end of such a run are illustrated in
Fig.~\ref{snap_0.01}. Two snapshots of the gas ({\it left panel})
and dust ({\it right panel}) density in the $(R,Z)$ plane are
shown in the case $(\Omega \tau_s)_0=0.01$ at time $t=600$. The dust
disc appears thinner than the gas disc, indicating that
significant settling has occurred.

In the following subsections, we will compare the dust distributions
we obtained for the different sizes to the models described in
Sect.~\ref{dust_settling_sec}. To do so, all relevant physical
quantities will be averaged in time between $t=550$ and $t=600$ using
$50$ snapshots so
that they become statistically significant. We first start by 
describing the relevant properties of the turbulence that result from
this procedure before concentrating on the degree of settling as a
function of size.

\subsection{Turbulence properties}
\label{turb_prop_sec}

\begin{figure}
\begin{center}
\includegraphics[scale=0.45]{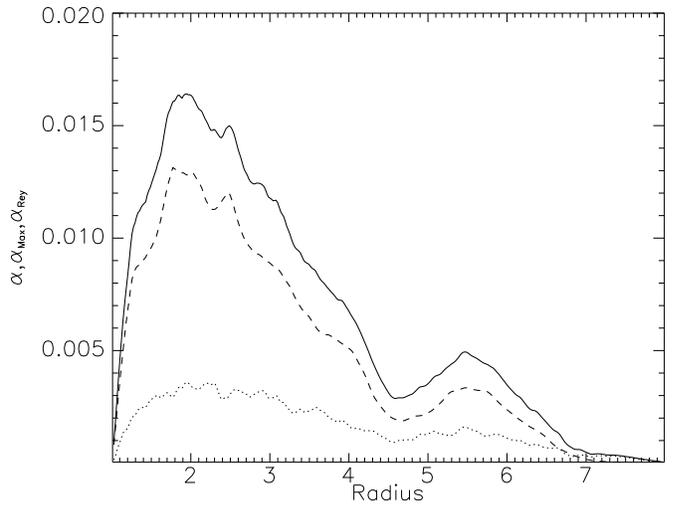}
\caption{Radial profile of $\alpha$ ({\it solid line}), $\alpha_{Max}$ 
({\it dashed line}) and $\alpha_{Rey}$ ({\it dotted line}) for the
  case $(\Omega \tau)_0=0.001$. The data have been averaged in time 
between $t=550$ and $t=600$.}
\label{alpha_mean_0.001}
\end{center}
\end{figure}

\begin{figure}
\begin{center}
\includegraphics[scale=0.45]{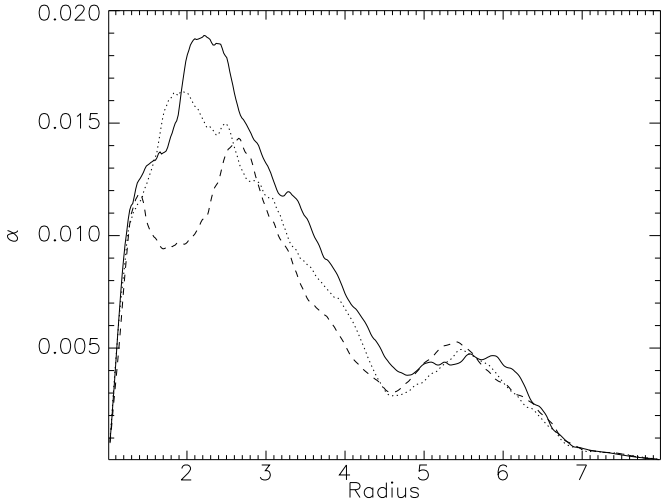}
\caption{Radial profile of $\alpha$ (time averaged between $t=550$ and 
$t=600$) for the models  $(\Omega \tau)_0=0.01$ ({\it solid line}),
  $0.001$ ({\it dotted line}) and $0.0001$ ({\it dashed
    line}). Angular momentum transport is similar at all radii in the
  three models, despite differences arising because of the stochastic
  nature of MHD turbulence (see text for details).}
\label{compar_alpha}
\end{center}
\end{figure}

\begin{figure}
\begin{center}
\includegraphics[scale=0.5]{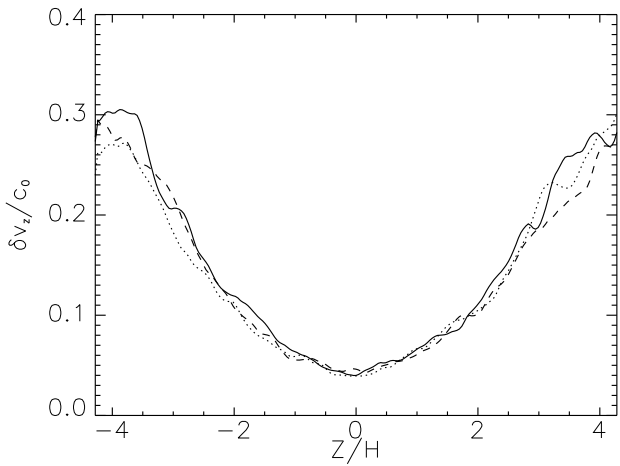}
\caption{Vertical profile of the vertical velocity fluctuations (in units of
  the speed of sound) for the model having $(\Omega \tau)_0=0.01$
  ({\it solid line}), $0.001$ ({\it dotted line}) and $0.0001$ ({\it
    dashed line}) at $R=2.93$. The simulations data have been averaged
  between $t=550$ and $t=600$.}
\label{vel_fluc}
\end{center}
\end{figure}

\begin{figure}
\begin{center}
\includegraphics[scale=0.45]{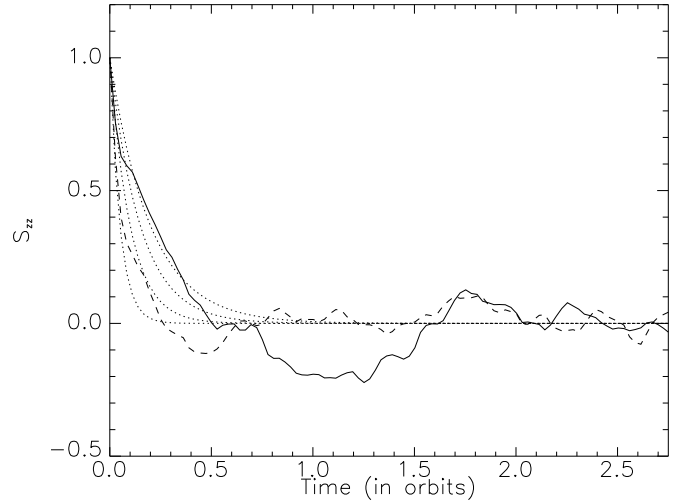}
\caption{Time variation of the correlation function of the vertical
  velocity fluctuations in the disc midplane ({\it solid line}) and in
the disc corona ({\it dashed line}). The dotted lines represent
functions decreasing exponentially toward zero with typical times
$\tau_0=0.05$,$0.1$, $0.15$ and $0.2$ orbit. They can be used to estimate
the typical correlation timescale of the turbulence. At both
locations, it is not far from the value $\tau_{corr}=0.15$ orbit we
used when modeling the results of the numerical simulations.}
\label{tau_corr_fig}
\end{center}
\end{figure}

The first relevant property of the turbulence is the vertically
averaged angular momentum transport it generates. It is
commonly measured using the parameter $\alpha$. Following
\citet{fromang&nelson06}, we calculated $\alpha$ as a function of
radius according to
\begin{equation}
\alpha=\alpha_{Rey}+\alpha_{Max}=
\Sigma \frac{\overline{ \delta v_R \delta v_{\phi}}-\overline{
   \frac{ B_R B_{\phi}}{4 \pi \rho}  }}{\overline{P}} \, ,
\label{alpha_eq}
\end{equation}
where $\alpha_{Rey}$ and $\alpha_{Max}$ correspond respectively
to the Reynolds and Maxwell stress contributions to $\alpha$.
The overbar symbols denote
density-weighted azimuthal and vertical averages 
\citep[see Eq.~(6) of][]{fromang&nelson06}. To reduce statistical
noise, $\alpha$ was further averaged in time between $t=550$ and
$t=600$. Its radial profile is shown in Fig.~\ref{alpha_mean_0.001}
for the model having $(\Omega 
\tau)_0=0.001$. The dashed and dotted lines, respectively, show the
variations of $\alpha_{Max}$ and $\alpha_{Rey}$, while the solid line
represents $\alpha$, the sum of the two. As described in
\citet{fromang&nelson06}, $\alpha$ presents large oscillations in
space and time, but its time averaged value is well behaved, varying
between $\sim 5 \times 10^{-3}$ and $\sim 1.5 \times 10^{-2}$, in
agreement with the results of \citet{fromang&nelson06}. As is
usually obtained in numerical simulations of this type, the Maxwell
stress dominates over the Reynolds stress by a factor of 
about two to four. The volume averaged value of $\alpha$ is also
  in agreement with the results of \citet{fromang&nelson06}. Indeed,
  we obtained $\alpha=5.6 \times 10^{-3}$, $5.5 
  \times 10^{-3}$, $7.5 \times 10^{-3}$, $7.1 \times 10^{-3}$, $6.4
  \times 10^{-3}$ and $6.6 \times 10^{-3}$ respectively at times
  $t=550$, $560$, $570$, $580$, $590$ and $600$. For the
  duration of the simulation, the turbulence is clearly in a
  quasi steady state.

In this paper, we present three simulations for different
particle sizes. Because these simulations were obtained under
different computing set-ups (i.e. using different numbers of
CPUs), the details of the turbulent flows
differ from one run to another. This is simply due to the chaotic
nature of turbulence \citep{wintersetal03}. 
However, the statistical properties of the
turbulence are similar. This is shown on Fig.~\ref{compar_alpha}
where we compare the radial profile of $\alpha$ for the models $(\Omega
\tau)_0=0.01$ ({\it solid line}), $0.001$ ({\it dotted line}) and
$0.0001$ ({\it dashed line}). For each case, the curves are averaged
in time between $t=550$ and $t=600$. Fig.~\ref{compar_alpha}
demonstrates that we obtain very similar values of $\alpha$ in the
different simulations. Thus it is meaningful to compare the dust
distributions in the three different models.

As explained in Sect.~\ref{varying_D}, the vertical velocity
fluctuations and the correlation timescale of the turbulence are also
of importance in affecting the diffusion of solid particles. In
Fig.~\ref{vel_fluc}, we show the vertical profile of the
vertical velocity fluctuations at $R=2.93$, normalized by the speed of
sound and averaged in time between $t=550$ and $t=600$ for the
models $(\Omega \tau)_0=0.01$ ({\it solid line}), $0.001$ ({\it dotted
line}) and $0.0001$ ({\it dashed line}). Again the results are very
similar for each model and in agreement with those of
\citet{fromang&nelson06}: the average 
$\delta v_z$ is of order $5 \%$ of the speed
of sound in the disc midplane before rising up to values of the order
of 20 -- 30\% in the disc corona, where weak shocks develop
in locations where convergent flow speeds exceed the sound speed
\citep{fromang&nelson06}.
These supersonic turbulent motions are driven by magnetic
stresses in regions where the Alfv\'en speed exceeds the sound speed.
Fig.~\ref{vel_fluc} shows that $\delta v_z$ varies by a
factor of about $5$ between the disc midplane and its corona.

As expressed by Eq.~(\ref{D_vel}), the dust diffusion coefficient also
depends on the turbulence correlation timescale
$\tau_{corr}$. Although $D$ depends on 
the value of $\tau_{corr}$ to the first power only,
while it depends on the velocity fluctuations to the
second power, vertical variations in the correlation timescale could
still affect the numerical estimate of the diffusion
coefficient. We thus calculated the value of $\tau_{corr}$ at two
locations, one in the disc midplane and the other in the disc
corona. Following \citet{fromang&pap06}, $\tau_{corr}$ 
can be evaluated by monitoring the time variation of the function
\begin{equation}
\textrm{S}_{\textrm{zz}}(\tau)= < v_z(z,t) v_z(z,t+\tau) > \, ,
\end{equation}
where $<.>$ denotes an ensemble average. $\textrm{S}_{\textrm{zz}}(\tau)$ is
expected to decrease toward zero from initially positive values in a
time $\tau_{corr}$. To estimate the later, the model in which $(\Omega
\tau)_0=0.01$ was restarted at time $t_0=573$. We then 
calculated and averaged the function $v_z(z,t_0) v_z(z,t_0+\tau)$ at
seven different radii $R=2$, $2.5$, $3$, $3.5$, 
$4$, $4.5$ and $5$ over a kernel of five cells in the radial
direction. At all radii, the function was further averaged over two
ranges in the meridional direction: 
$|\theta| \le 1.5 H/R$ to produce the function
$\textrm{S}_{\textrm{zz1}}$, and $|\theta| \in [1.5 H/R,3.5 H/R]$ 
to produce the function $\textrm{S}_{\textrm{zz2}}$.
$\textrm{S}_{\textrm{zz1}}$ thus represents the
velocity correlation function near the disc midplane while
$\textrm{S}_{\textrm{zz2}}$ represents the correlation function in the
upper layers including the corona. The two functions are 
plotted as a function of
$\tau$ in Fig.~\ref{tau_corr_fig}, respectively, with a solid and a
dashed line (both curves are normalized by their value at
$\tau=0$). As expected, both display an initial decrease toward zero
around which 
they stabilize after a few tenths of an orbit. This qualitative trend is
in agreement with the results of \citet{fromang&pap06}. The dotted curves
over plotted on the same figure represent the functions
$\textrm{S}_{\textrm{zz}}^{\tau_0}=\textrm{exp}(-t/\tau_0)$ for 
$\tau_0=0.05$, $0.1$, $0.15$ and $0.2$ orbits. Although it is
difficult to measure the correlation timescale precisely given the
large fluctuations we obtained, these curves can still be used as a
way to estimate the correlation time $\tau_{corr}$ governing the
functions $\textrm{S}_{\textrm{zz1}}$ and
$\textrm{S}_{\textrm{zz2}}$. They indicate that $\tau_{corr}$ is $\sim
0.05-0.1$ in the corona and $\sim 0.1-0.2$ in the disc midplane. Both
values are comparable to the value of $0.15$ orbit reported by
\citet{fromang&pap06}. Although these two timescales are different,
their ratio is at most two
and certainly accounts for less variation in the diffusion
coefficient $D$ than the variations in the vertical velocity
fluctuations described above. Therefore, when plugging numerical
estimates for the correlation time into Eq.~(\ref{D_vel}), we will use
$\tau_{corr}=0.15$ orbits in the remaining of this paper. This
  value is such that $\Omega \tau_{corr} \sim 1$ and is in agreement
  with previously reported results in the literature
  \citep{johansenetal06} and with values used to model the effect of
  turbulence in theoretical studies of planet formation \citep[see for
  example][]{weidenschilling84}, thus giving additional support to such work.

\subsection{Dust spatial distribution}

\begin{figure*}
\begin{center}
\includegraphics[scale=0.29]{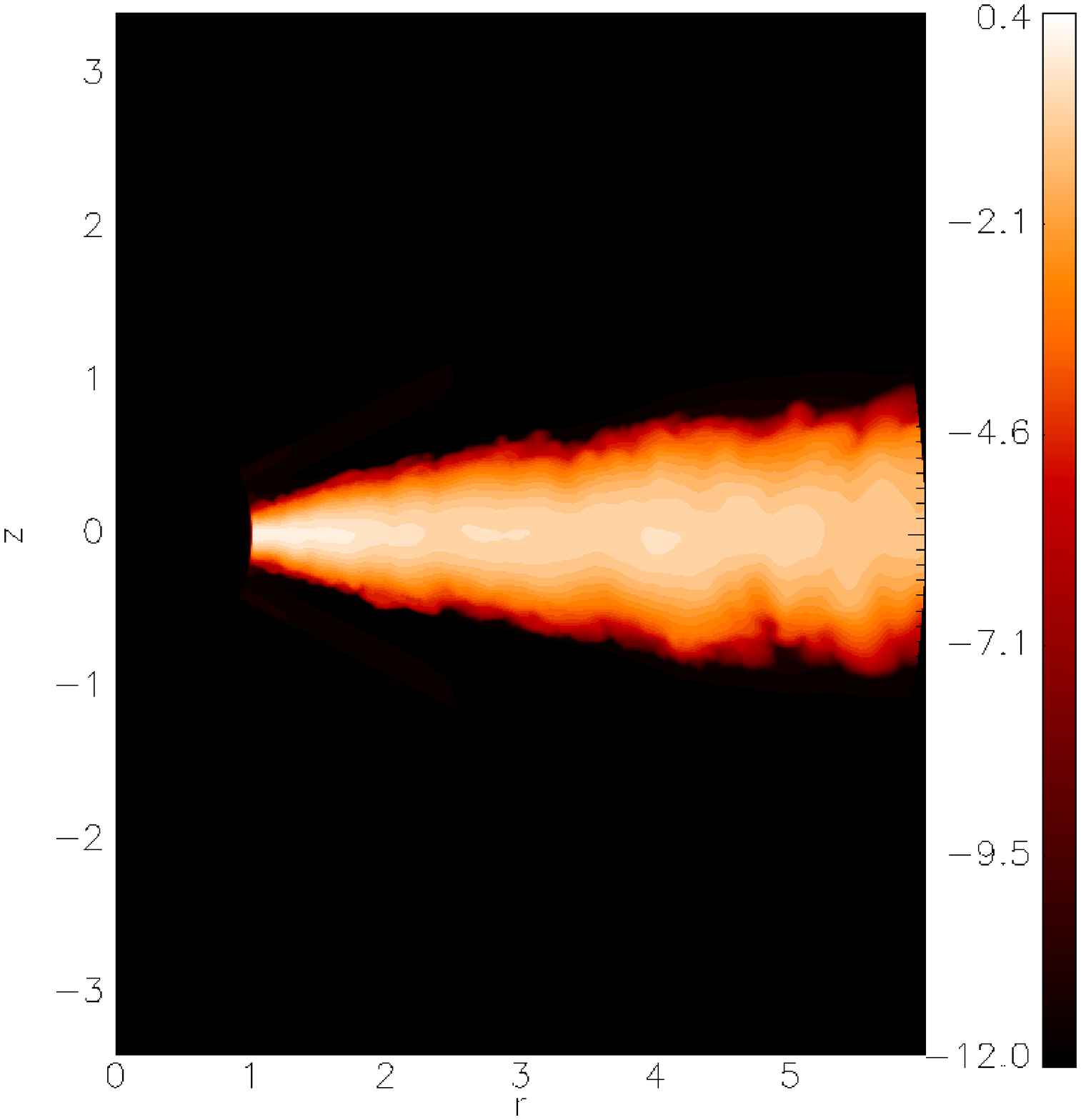}
\includegraphics[scale=0.29]{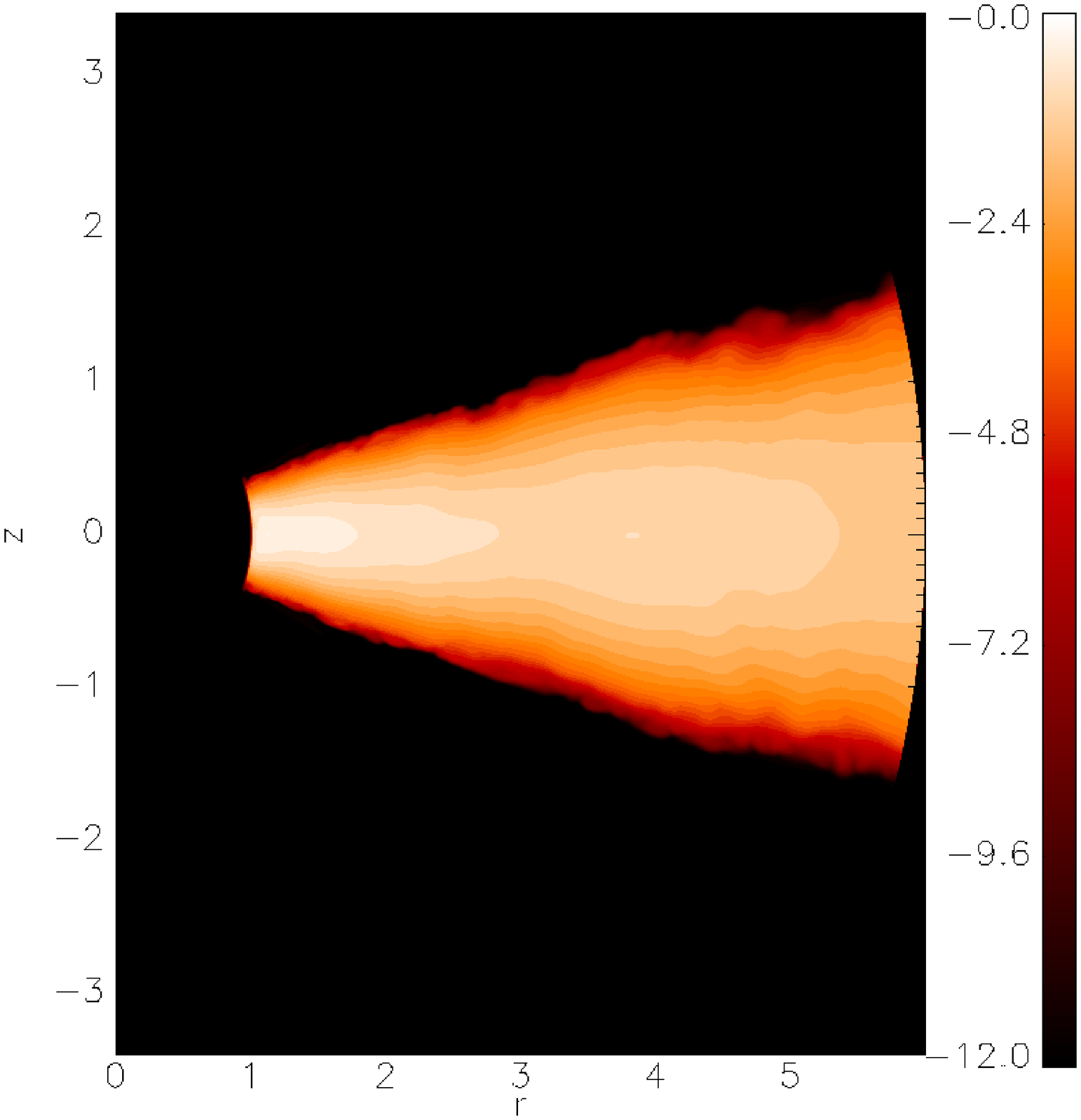}
\includegraphics[scale=0.29]{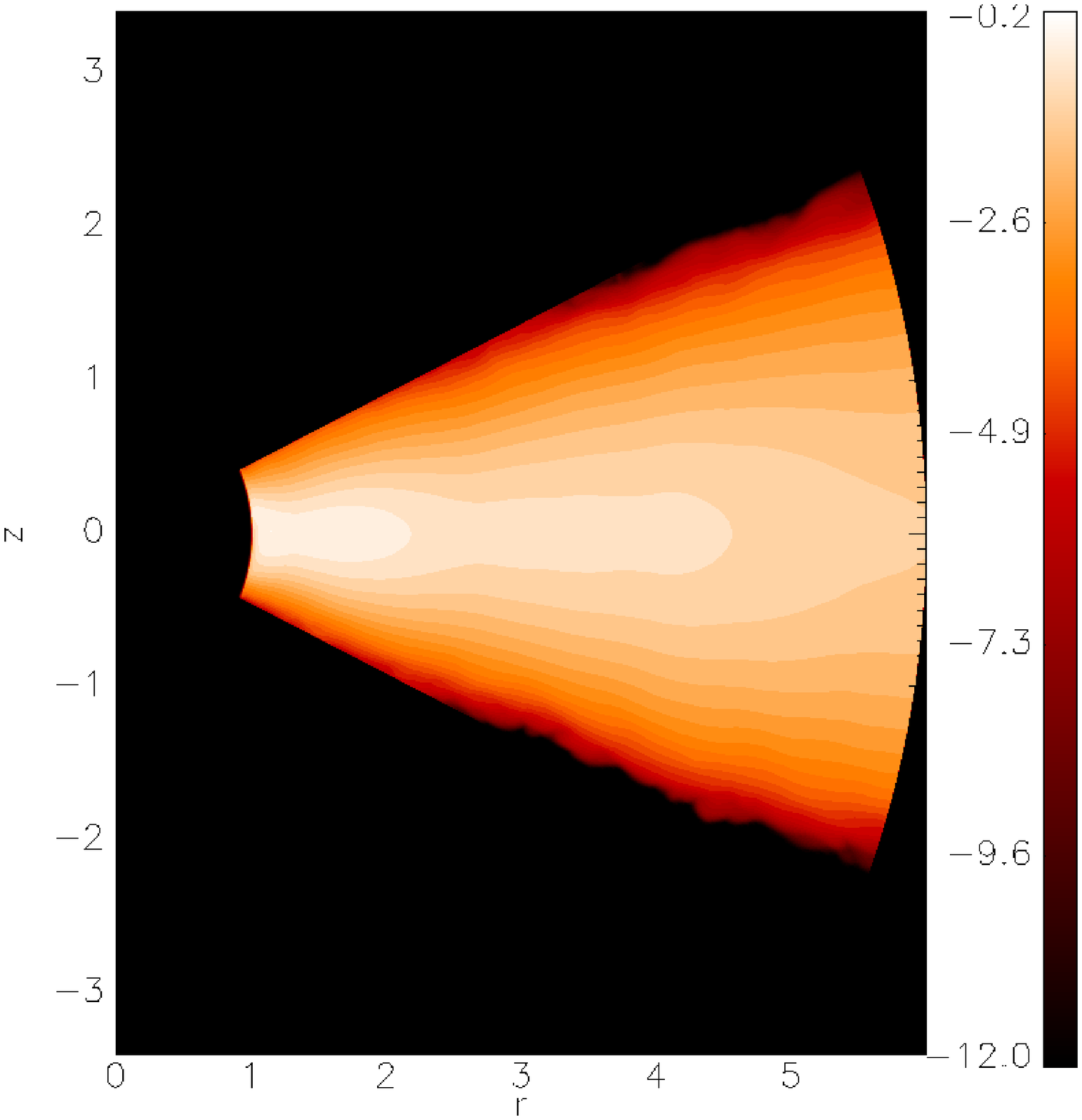}
\caption{Dust density distribution in the $(R,Z)$ plane in the numerical simulation 
 for the model having $(\Omega \tau)_0=0.01$ ({\it left panel}),
 $(\Omega \tau)_0=0.001$ ({\it middle panel}) and $(\Omega
 \tau)_0=0.0001$ ({\it right panel}). For all cases, the raw data have
 been first azimuthally averaged and then time averaged between $t=550$
 and $t=600$.}
\label{simu_dust_2d}
\end{center}
\end{figure*}

\begin{figure*}
\begin{center}
\includegraphics[scale=0.29]{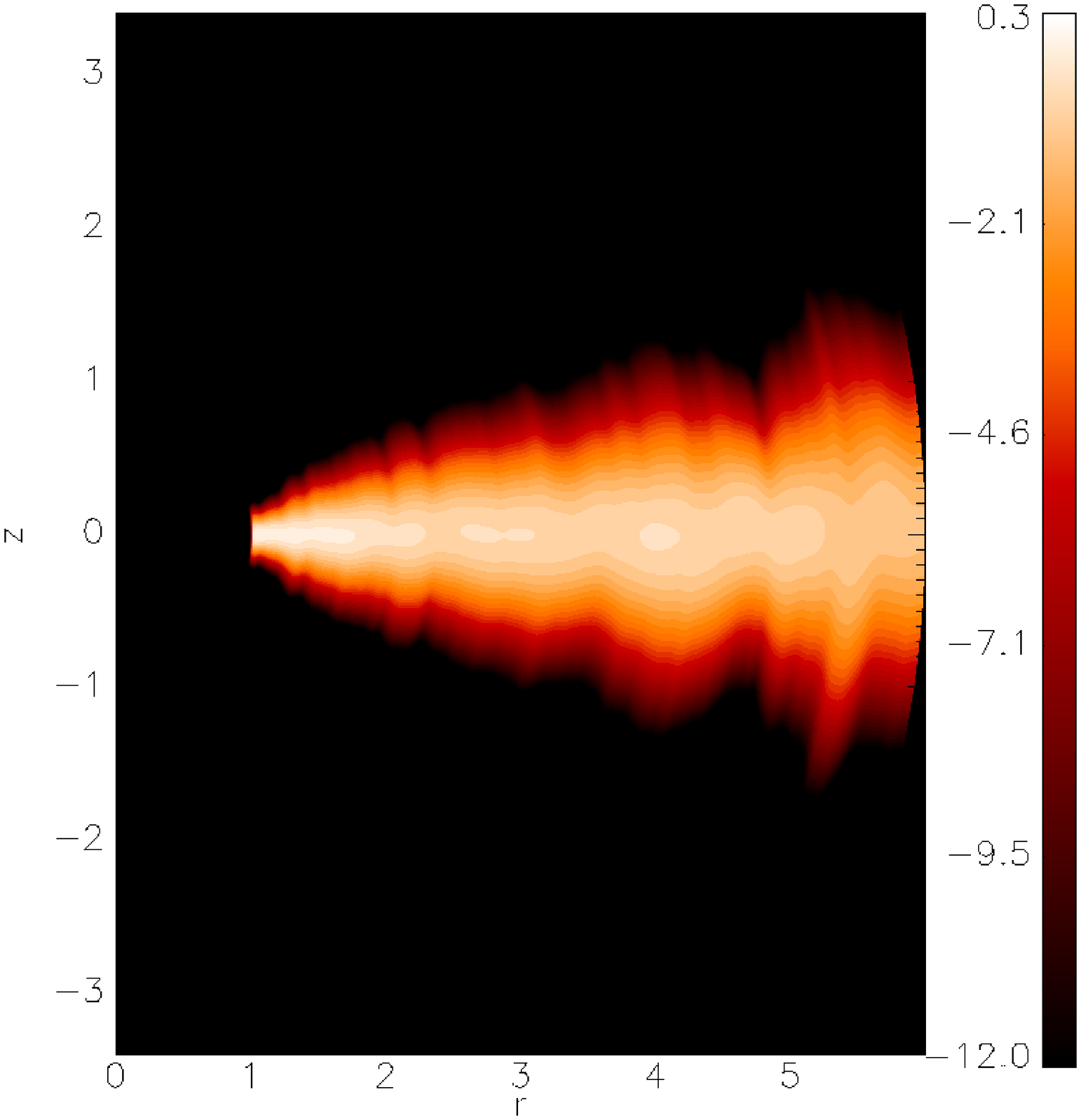}
\includegraphics[scale=0.29]{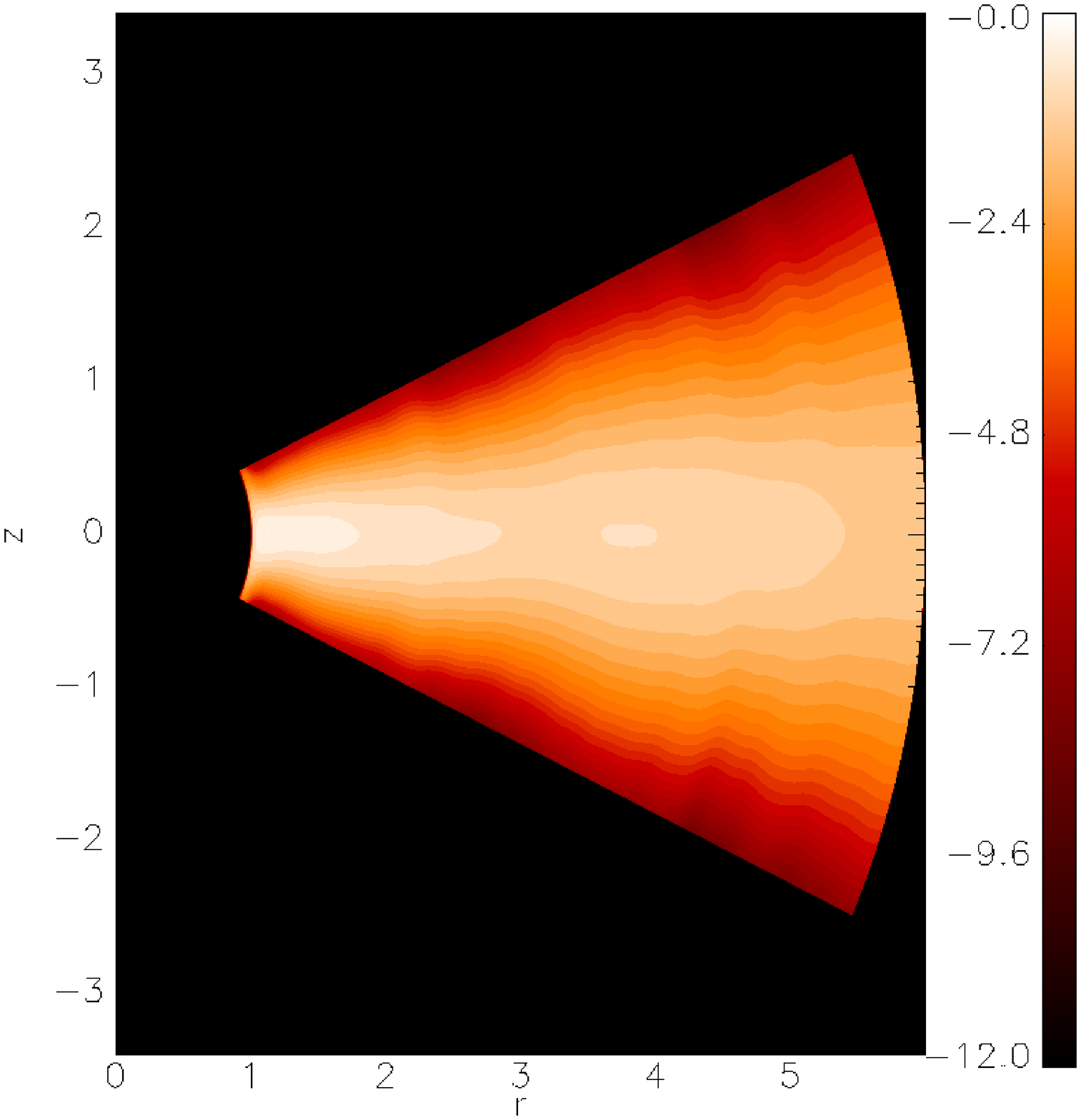}
\includegraphics[scale=0.29]{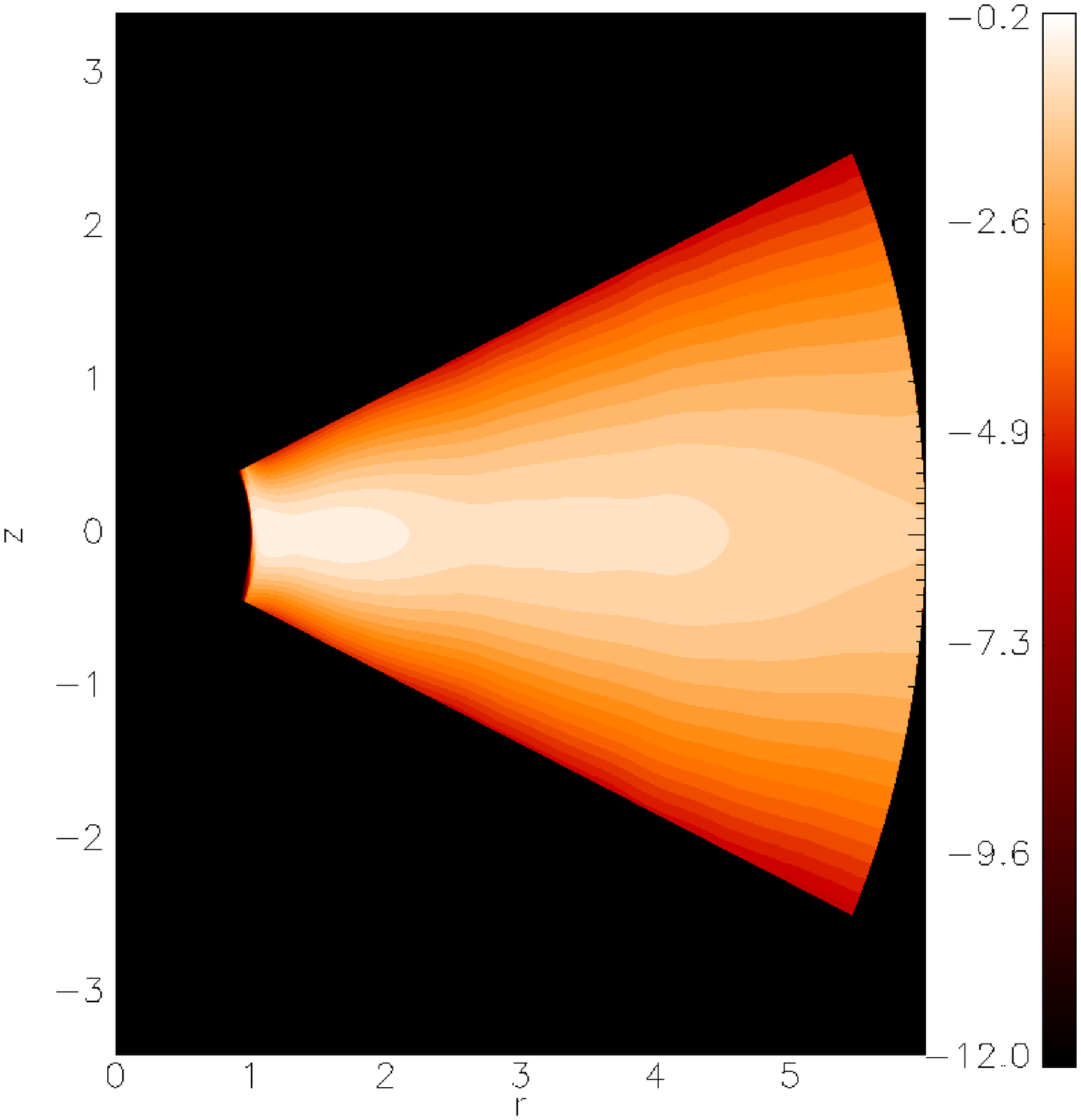}
\caption{Dust density distribution in the $(R,Z)$ plane computed
  by fitting a Gaussian vertical profile to the simulation data
  (azimuthally and time averaged) at each radii. As is the case for
  fig.~\ref{simu_dust_2d}, the left, middle and right panels
  respectively correspond to $(\Omega \tau)_0=0.01$, $0.001$ and $0.0001$.}
\label{dust_gauss}
\end{center}
\end{figure*}

\begin{figure*}
\begin{center}
\includegraphics[scale=0.3]{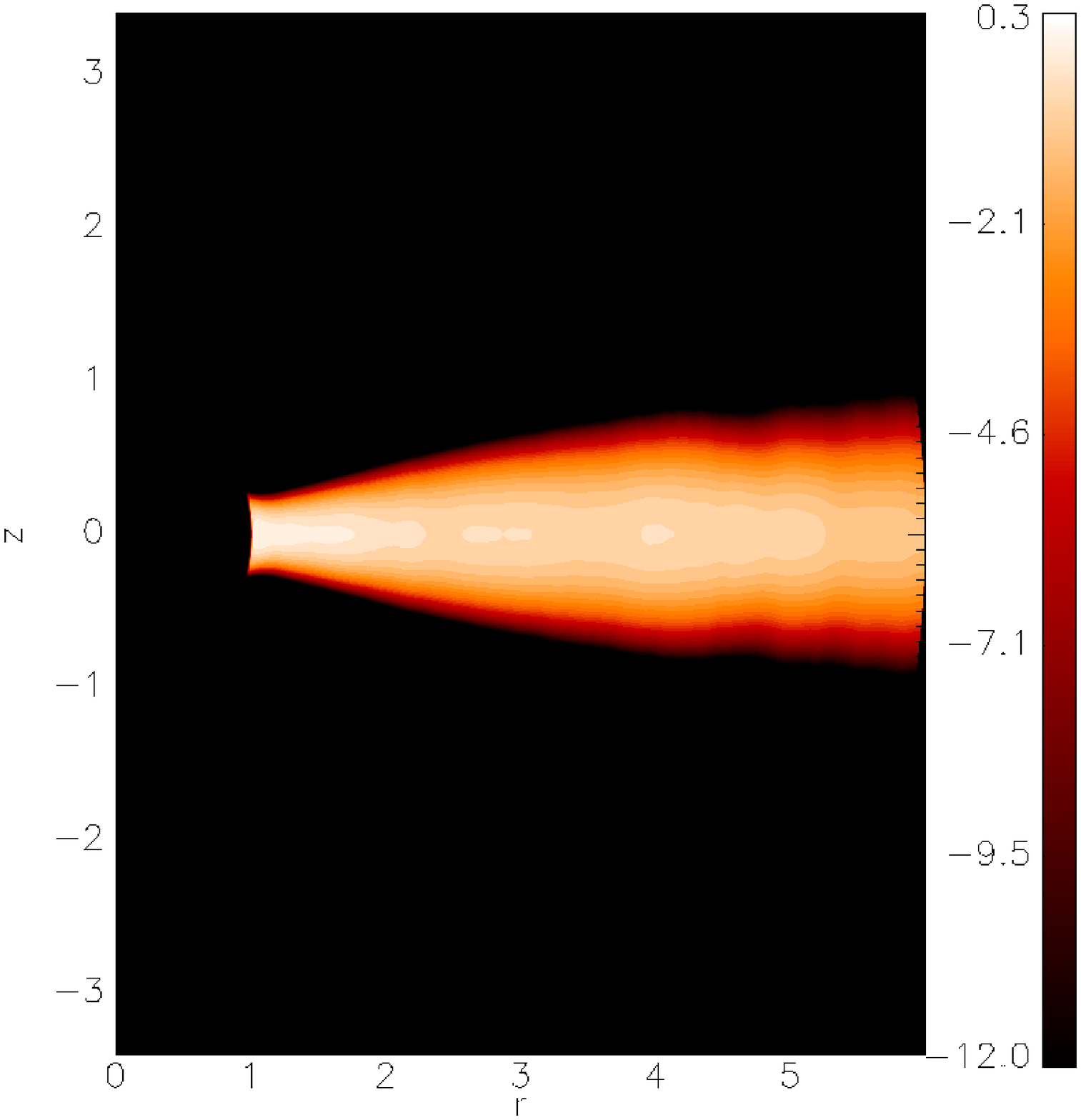}
\includegraphics[scale=0.3]{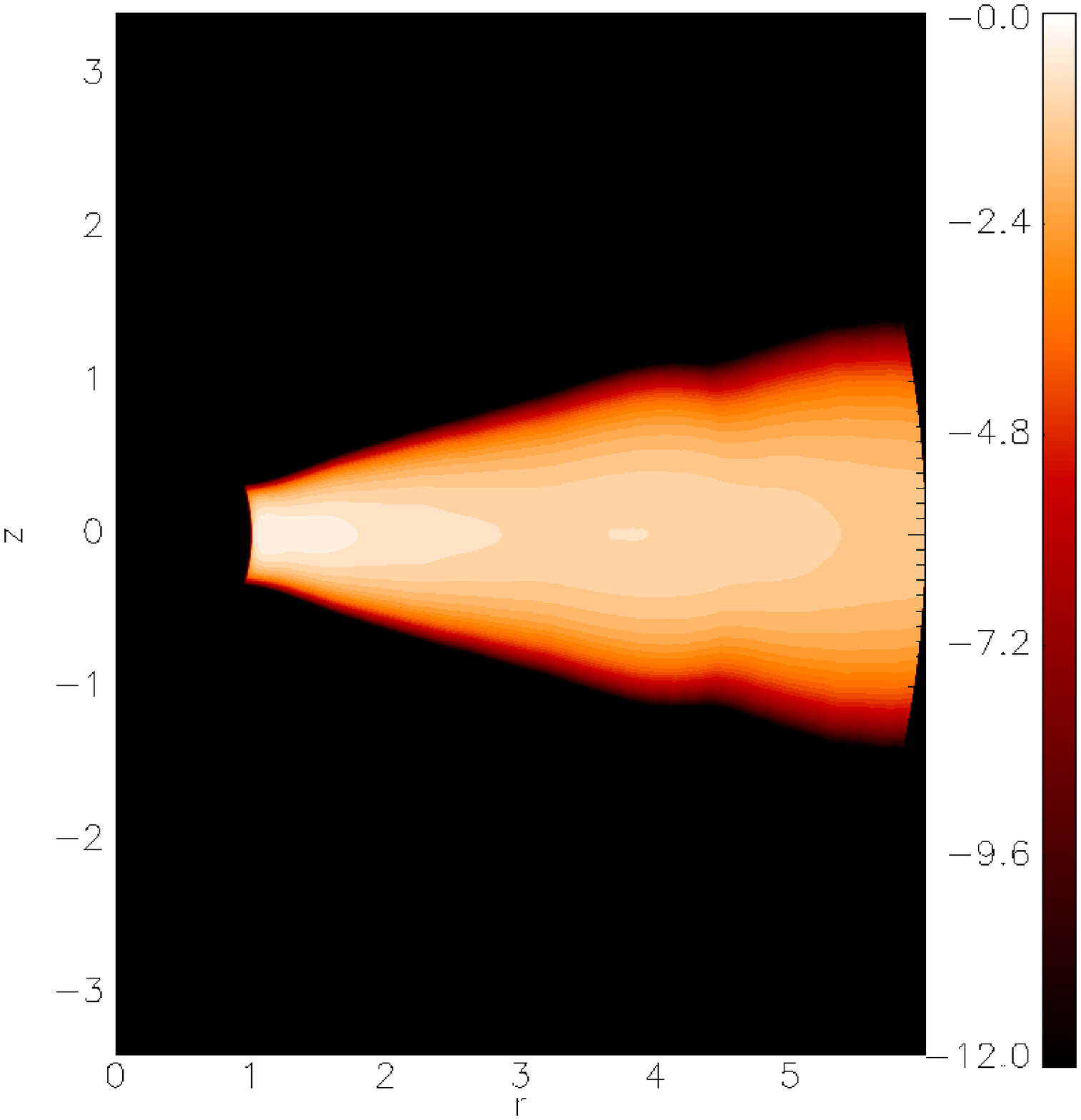}
\includegraphics[scale=0.3]{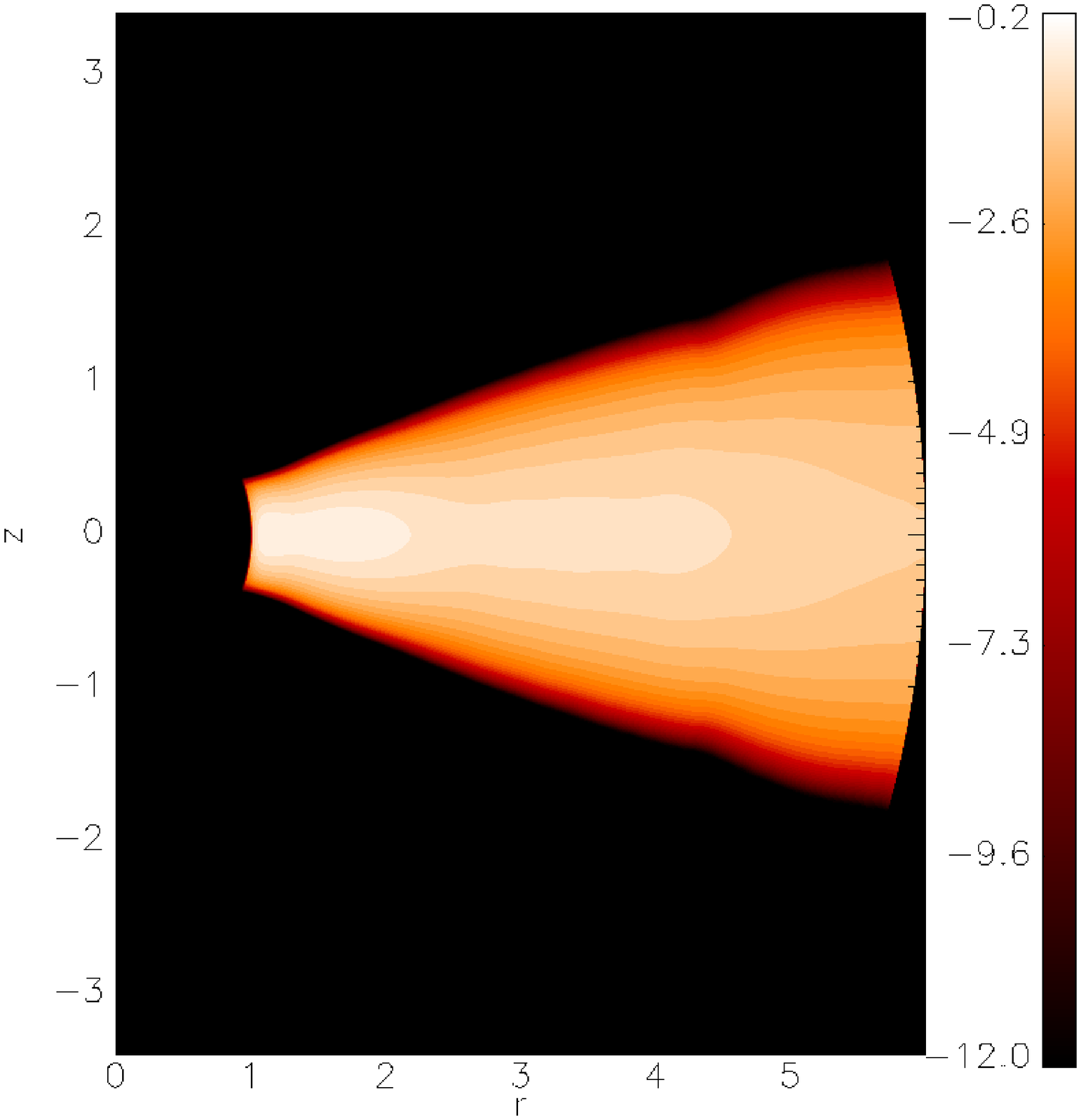}
\caption{Same as fig.~\ref{dust_gauss} using the model that 
assumes $D=\textrm{constant}$.}
\label{dust_Dconst}
\end{center}
\end{figure*}

\begin{figure*}
\begin{center}
\includegraphics[scale=0.29]{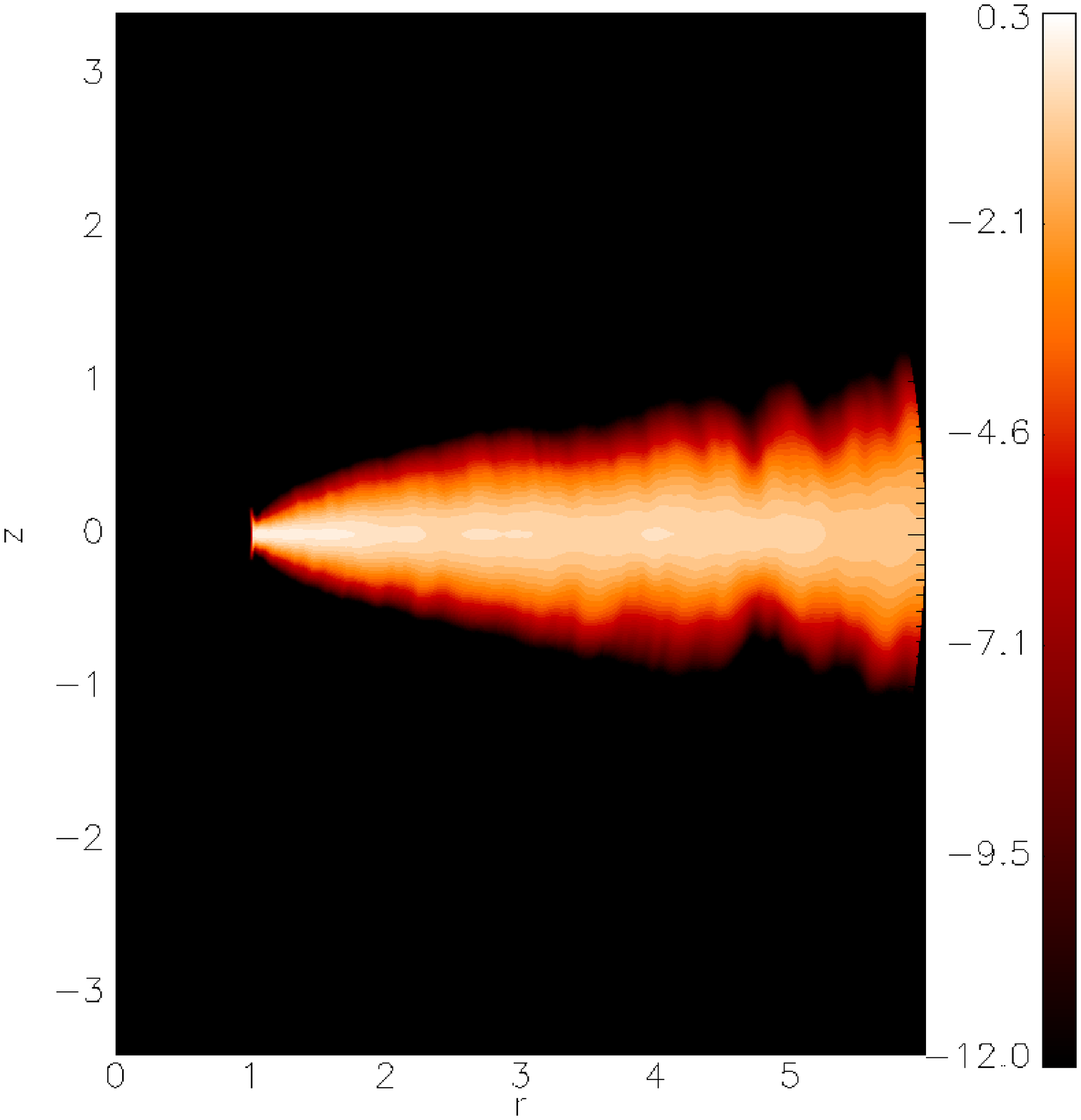}
\includegraphics[scale=0.29]{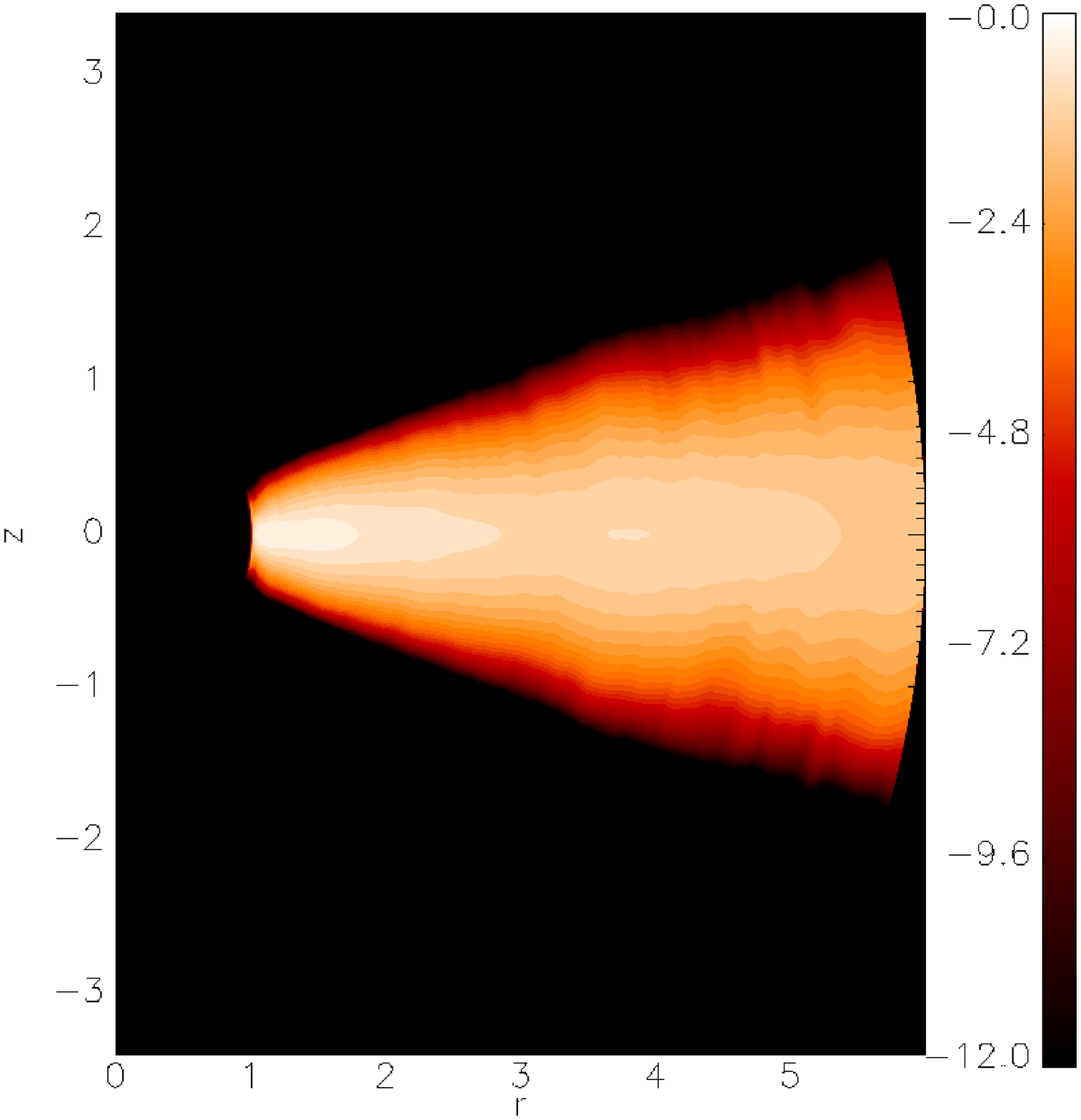}
\includegraphics[scale=0.29]{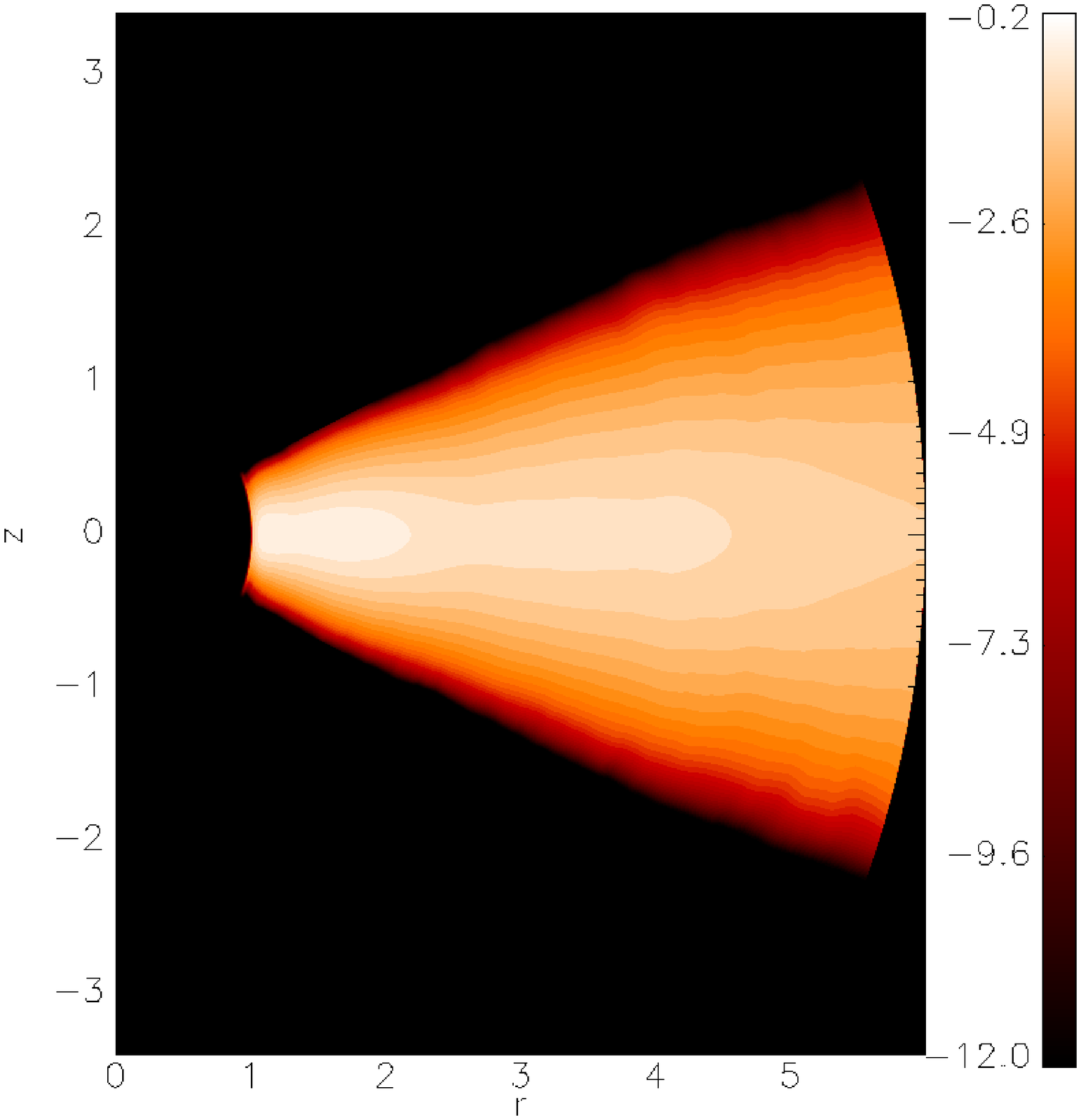}
\caption{Same as fig.~\ref{dust_gauss} using the model that assumes
  $D=\delta v_z^2 \tau_{corr}$.}
\label{dust_Dvar}
\end{center}
\end{figure*}

For the three simulations we performed, we averaged the dust density
in azimuth and over time between $t=550$ and $t=600$ orbits.
The spatial distributions we obtained using this procedure
are shown in Fig.~\ref{simu_dust_2d} for the models having $(\Omega
\tau)_0=0.01$ ({\it left panel}), $0.001$ ({\it middle panel}) and
$0.0001$ ({\it right panel}). Note that the radial extent of the
snapshots is limited to $R \leq 6$. At larger radii, the outer buffer
region we use \citep[see][]{fromang&nelson06} starts to affect the dust
distribution. As expected, the smaller the dust
particle sizes, the thicker the dust disc. This is simply because
smaller particles are better coupled to the gas and can thus be lifted
further away from the disc midplane by the turbulence before they
decouple from the gas. The left panel also shows some sign of the dust
disc flattening at large radii. This is simply because the strength
of the turbulence, as measured for example by the parameter $\alpha$, 
decreases at large radii, as seen in Figs.~\ref{alpha_mean_0.001} and
\ref{compar_alpha}. It should not be confused with the apparent 
dust disc flattening identified by \citet{dullemond&dominik04},
which occurs because of self--shadowing in the presence of 
weak turbulence. Finally, Fig.~\ref{simu_dust_2d} also highlights one
of the limitations of our work: with the set-up of the simulations
presented in this
paper, we cannot easily extend our parameter survey to smaller
particles. Indeed, when $(\Omega \tau)_0=0.0001$, the dust disc
already covers almost the entire computational domain
in the vertical direction. Reducing
further the size of the dust particles would require an increase in
the size of the computational domain in the meridional region for the
decoupling between gas and dust to occur within the computational
grid. This would require an increase in the computing time required
for a simulation performed with the same resolution. 
This will soon become possible as
computational resources improve, but is currently very
challenging. 

To compare these results with the models presented in
Sects.~\ref{gauss_prof_sec}, \ref{const_D} and \ref{varying_D}, we
computed the expected 2D dust distribution that each of them would
predict. In doing so, we used the same disc and dust parameters as in
the simulations. 
\begin{itemize}
\item[-] For the
first model (presented in Sect.~\ref{gauss_prof_sec}), we fitted a
Gaussian profile to the vertical profile of
the dust density at each radius. This was done by performing a least
  squares fit to the dust density profile over the entire vertical
  extent of the disc. The resulting spatial distributions
are plotted in Fig.~\ref{dust_gauss} using the same color table and
spatial domain as in Fig.~\ref{simu_dust_2d}. 
\item[-] For the second model
(presented in Sect.~\ref{const_D}), we used Eq.~(\ref{dust_prof_eq})
to calculate the vertical profile of the dust density at each
radius. The numerical value of the dimensionless diffusion coefficient
$\tilde{D}$ at each radius was calculated using 
Eq.~(\ref{const_D_eq}) in which we plugged the value of $\alpha$
calculated according to Eq.~(\ref{alpha_eq}). We used ${\rm Sc}=1.5$ as this
value turns out to provide the best fit to the simulations. Being of
order unity, it is also in agreement with the results of previous
local numerical simulations
\citep{johansen&klahr05,johansenetal06,fromang&pap06}. The resulting spatial
distributions of the dust density are plotted in
Fig.~\ref{dust_Dconst}. 
\item[-] For the third model (presented in
Sect.~\ref{varying_D}), we integrated Eq.~(\ref{diff_advec_eq_II})
numerically, using at each disc altitude the estimate for the
diffusion coefficient $D$ provided by Eq.~(\ref{D_vel}). In this
equation, we used the azimuthally and time averaged vertical profile of the
vertical velocity fluctuations at each radius (as shown on
Fig.~\ref{vel_fluc} for the special case $R=2.93$) and the
correlation time $\tau_{corr}=0.15$ orbit as explained in
Sect.~\ref{turb_prop_sec}. The resulting spatial
distributions of the dust density are plotted in
Fig.~\ref{dust_Dvar}.
\end{itemize}

The spatial distributions of the dust density obtained with
the three different models are in rough agreement with the simulations
and with naive expectations: all predict that the dust component
of the discs thicken as dust particles decrease in size,
in agreement with the MHD simulations. For the
model having $(\Omega \tau)_0=0.01$ (i.e. for the largest
particles), the agreement between the three models and the
simulation is quite good. This can be understood easily:
in this model, the dust scale-height is about $0.3H$. In other words,
the solid
particles concentrate close to the equatorial plane, where the
turbulence properties are fairly homogeneous (the vertical velocity
fluctuations only start to rise significantly above $\sim 1.5H$). Thus
the diffusion coefficient calculated according to
Eq.~(\ref{D_vel}) is roughly constant and the second and third models
(with constant and varying diffusion coefficient)
give a similar result. Moreover, close to the equatorial plane (i.e. $Z
\ll H$), the leading order expansion to Eq.~(\ref{dust_prof_eq}) turns
out to be Gaussian, which explains the similarities between the three
models for large particles.

On the other hand, models having smaller dust sizes, i.e. $(\Omega
\tau)_0=0.001$ and $0.0001$, show differences between the different
approaches. The Gaussian fit to the simulations always overestimates
the dust density in the disc corona ($Z/H \geq 2$),
showing that in general the vertical profile of the dust density
distribution is non Gaussian. On the other hand, the model
having a constant diffusion coefficient always underestimates 
the dust density in the disc corona. Only the model taking the
vertical variation of the diffusion coefficient into account gives 
a satisfactory fit to the simulations, especially in the disc upper
layers.  

\begin{figure}
\begin{center}
\includegraphics[scale=0.5]{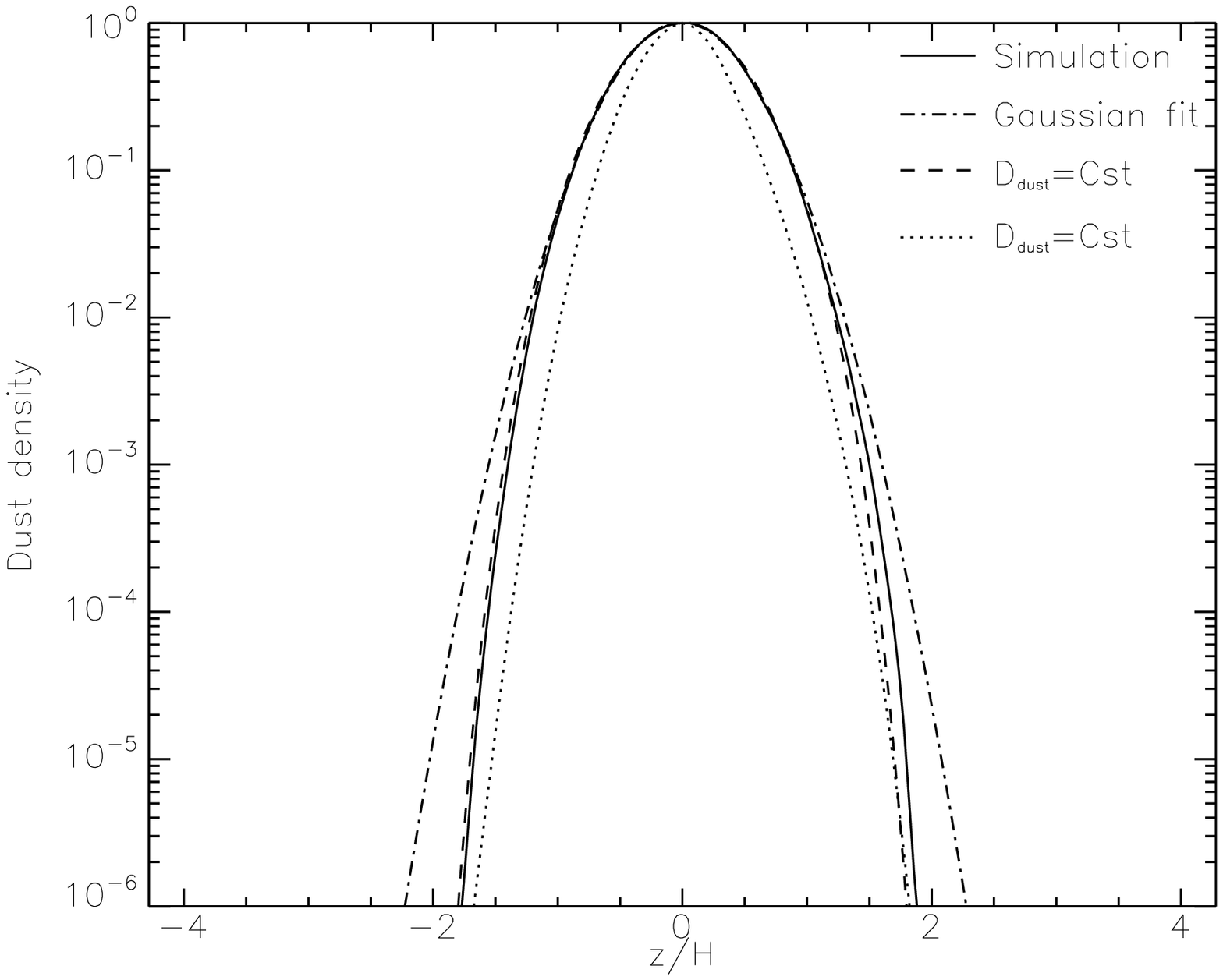}
\includegraphics[scale=0.5]{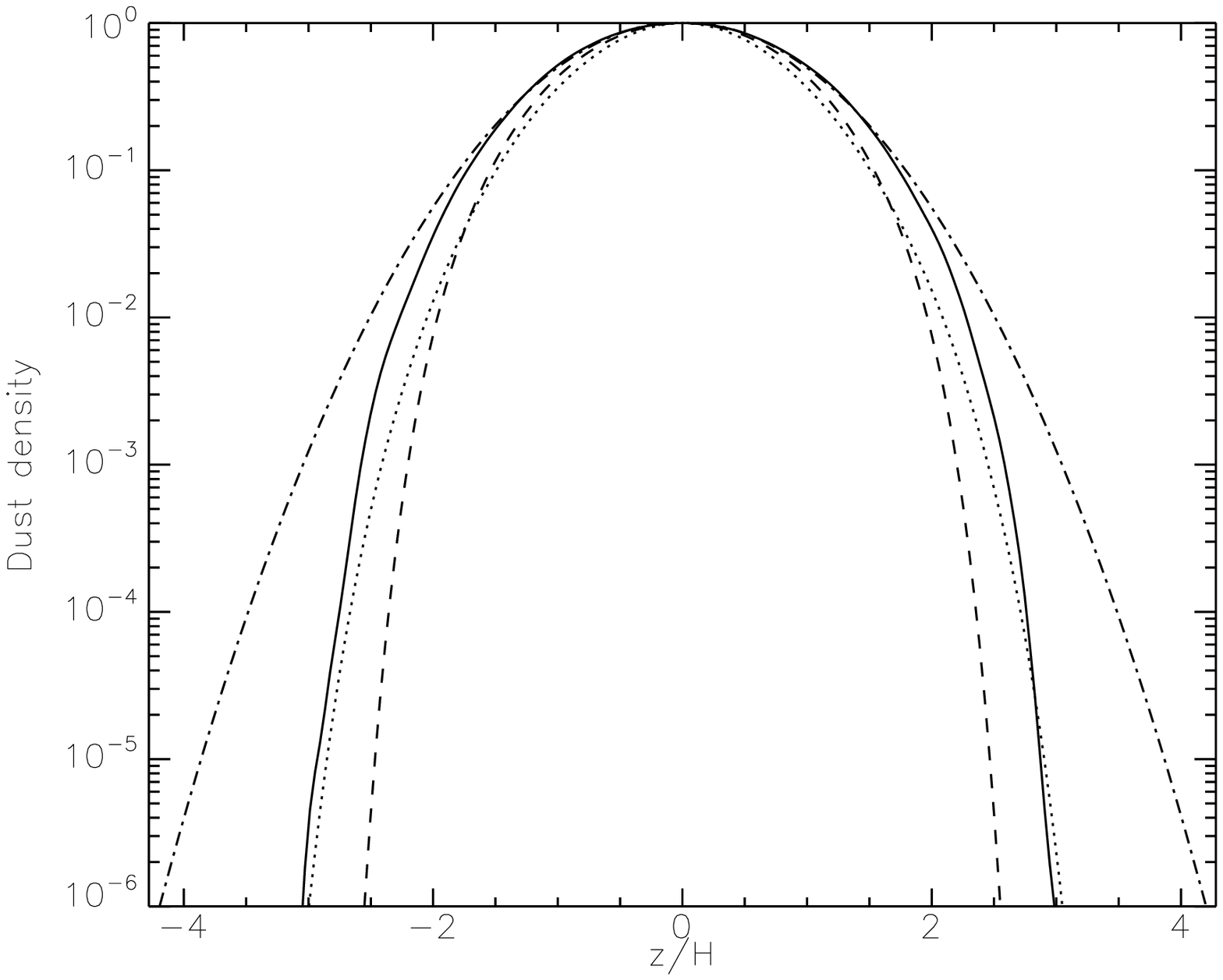}
\includegraphics[scale=0.5]{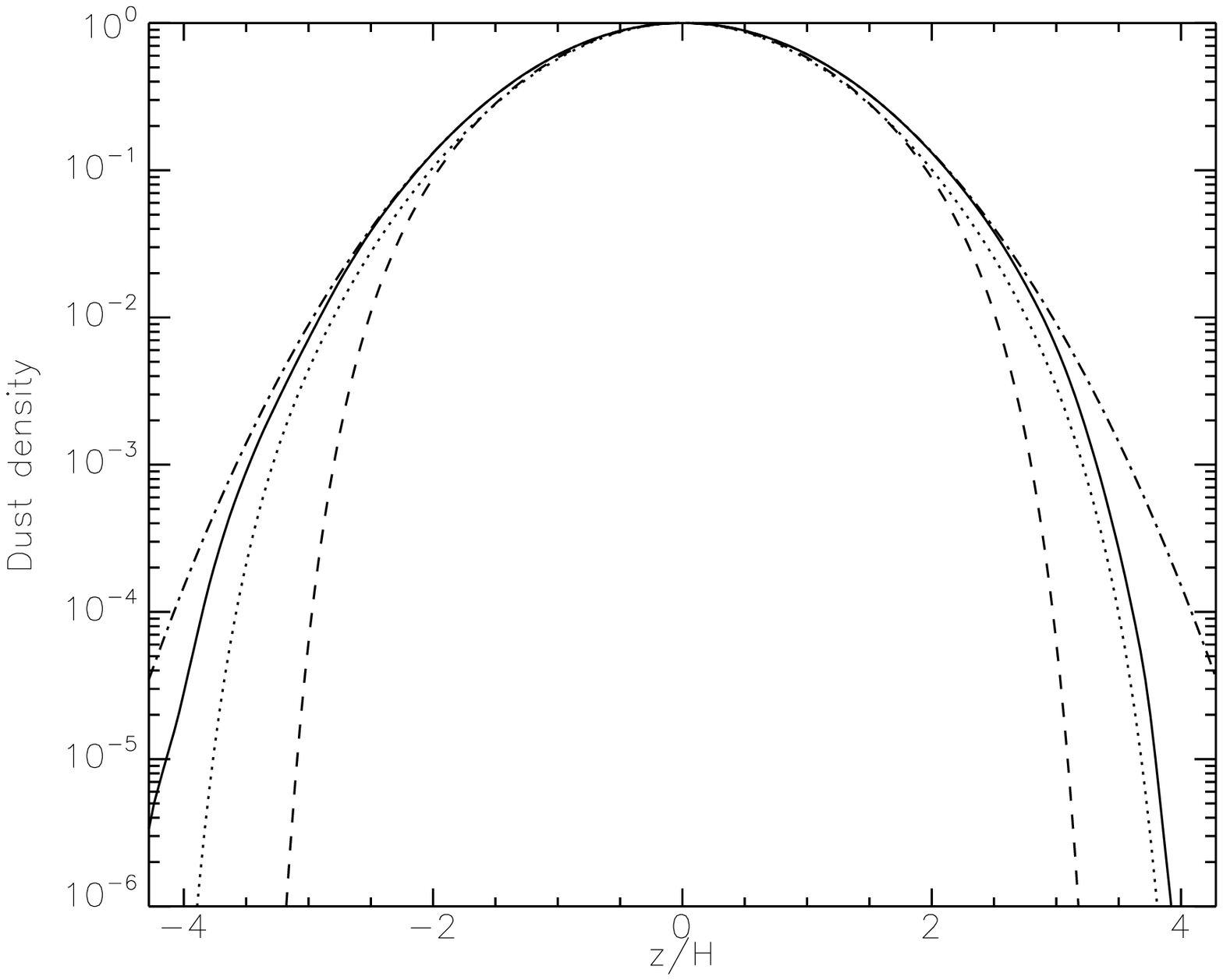}
\caption{Vertical profiles of the dust density, radially averaged
  between $R=3$ and $R=5$. As shown by the legend displayed on
    the upper panel, the different curves (normalized by their
  midplane values) were obtained using 
  the numerical simulations ({\it solid line}), by fitting a Gaussian profile
  to the simulation data ({\it dotted--dashed line}), using a constant
  turbulent diffusion coefficient ({\it dashed line}) and a vertically
  varying diffusion coefficient ({\it dotted line}). The upper, middle
  and bottom panels
  respectively correspond to $(\Omega \tau)_0=0.01$, $0.001$ and
  $0.0001$. In the first case, all models successfully reproduced the
  simulation, while for the smaller dust sizes, only the later and more
  elaborate model produces satisfactory results.}
\label{dust_prof}
\end{center}
\end{figure}

These differences can be made more quantitative by comparing the
vertical profile of the dust density between the simulations and the
models. This is done in
Fig.~\ref{dust_prof}. The left panel gathers the results corresponding
to the case $(\Omega\tau)_0=0.01$, the middle panel shows results
obtained when $(\Omega\tau)_0=0.001$. Finally, the right panel shows
results obtained when $(\Omega\tau)_0=0.0001$. In each panel, the
solid curves plot the vertical profile of the dust density obtained in
the MHD simulation by temporally, azimuthally and radially 
averaging the results.
The radial averaging is performed over the range $R \in [3,5]$. The
dot--dashed, dashed and dotted curves, respectively, correspond to
the model using a Gaussian fit to the data, a constant diffusion
coefficient and a vertically varying diffusion coefficient. Again, the
left panel demonstrates the good agreement between all approaches in
the case of large particles, as the curves used to represent the
results of the different models are almost undistiguishable on that
plot. It also provides, however, a first hint that the Gaussian
fit increasingly overestimates the dust density as one moves
away from the midplane, even for these larger particles. 
The middle and right panels confirm these results and 
show that the model having a vertically varying diffusion
coefficient provides in general the best fit to the data in the disc
corona. For example, in the case $(\Omega\tau)_0=0.001$, the three
models give a good fit to the data for $Z \leq 2H$. For $Z \geq 2H$,
the Gaussian fit to the data overestimates the density obtained in the
simulations (their ratio is about $10^3$ at $Z=3H$) and the model
having a constant diffusion coefficient underestimates that density
(the ratio at $3H$ is in excess of $10^6$). The agreement between the
simulations and the model having a vertically varying diffusion
coefficient is better as the ratio between the predicted and observed
dust densities is
always bounded by $10$, despite the fact that the dust density
itself varies by more than $6$ orders of magnitude. It is interesting to
point out, though, that the value of the density predicted by this
model close to the equatorial plane seems to significantly
underestimate the results of the simulations. This is most likely 
because the correlation timescale we used in calculating the diffusion
coefficient underestimates the midplane correlation timescale of the
turbulence (see section~\ref{turb_prop_sec}). Using a varying
correlation timescale would certainly 
further improve the agreement with the numerical simulations, but such
a level of refinement is probably meaningless given the other
approximations involved in the simulations themselves. 
As shown on Fig.~\ref{dust_prof}, the situation is similar in the case
$(\Omega\tau)_0=0.0001$: the best fit to the numerical simulations is
provided by the final and more elaborate model. Indeed, at $Z \geq
2.5H$, the dust density predicted by the model with a constant
diffusion coefficient starts to underestimate significantly the
results of the numerical simulations. Above $Z=3H$, the difference
becomes enormous, as in the case $(\Omega\tau)_0=0.001$. The
agreement, however, seems better with the model that uses a Gaussian
fit to the data than in the case $(\Omega\tau)_0=0.001$. The Gaussian
fit indeed gives a good fit up to $Z=3H$. This apparent agreement
would most probably break 
down for $Z \geq 4.3$, as the ratio with the simulated densities is
already greater than two orders of magnitude at $Z=4H$ and increases 
with increasing $Z$.

\section{Discussion and conclusions}
\label{conclusion_section}

In this paper, we have studied dust settling in turbulent
protoplanetary discs using global MHD numerical simulations performed
with the code GLOBAL using spherical coordinates. In this section, we
summarize our main findings and discuss the limitations and future
prospects of our work.

\subsection{Dust density vertical profile}

\begin{figure}
\begin{center}
\includegraphics[scale=0.5]{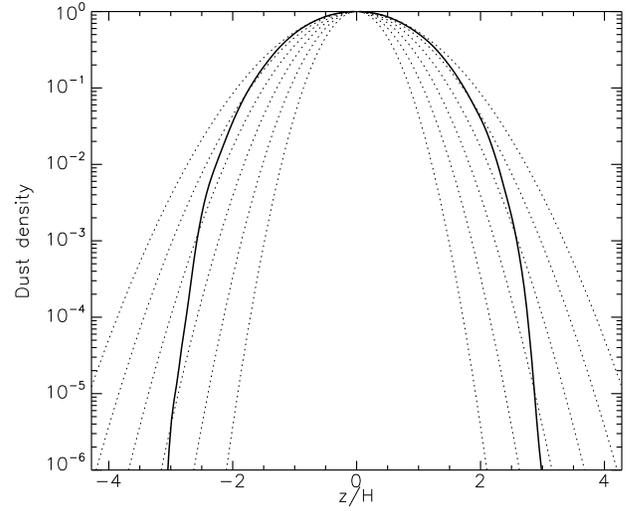}
\caption{This figure compares the dust density vertical profile (shown
  with a solid line), in the case $(\Omega\tau)_0=0.001$, with a set
  of Gaussian profiles with different scale-heights $H_{d,th}$ (shown with
  dotted lines). The different values of $H_{d,th}$ we used are $0.4H$,
  $0.5H$, $0.6H$, $0.7H$, $0.8H$ and $0.9H$. None of the dotted lines
  provides an acceptable fit to the numerical dust density. This
  demonstrates that the dust vertical profile is not Gaussian.}
\label{compar_gauss_0.001}
\end{center}
\end{figure}

\begin{figure}
\begin{center}
\includegraphics[scale=0.5]{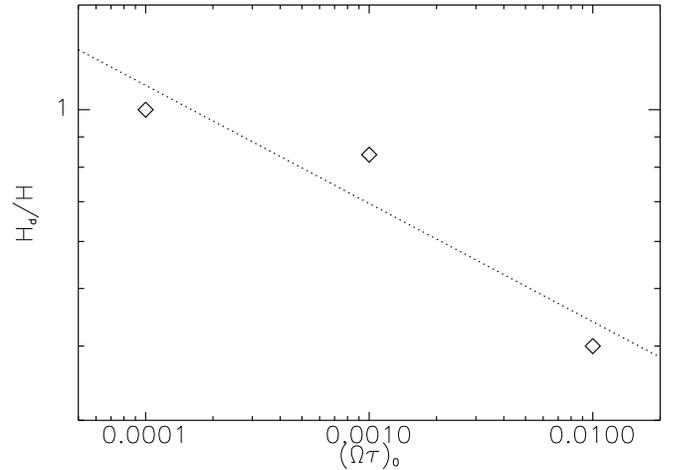}
\caption{Dust disc scale-heights obtained through a Gaussian fit to the
data are represented by the diamonds as a function of $(\Omega
\tau)_0$. The solid line displays the best power law fit to the points and
indicates that $H_d/H \propto a^{-0.2}$.}
\label{hd_fit}
\end{center}
\end{figure}

The first point that emerges is that Gaussian profiles
fail badly to reproduce the data extracted from the
simulations. This point is further illustrated by
Fig.~\ref{compar_gauss_0.001} where we compare the dust profile in
the case $(\Omega\tau)_0=0.001$ ({\it solid line}) with a
set of Gaussian profiles ({\it dotted lines}) computed according to
the following equation, 
\begin{equation}
\rho_{d,th}=\exp \left( -\frac{z^2}{2 H_{d,th}^2} \right) \, ,
\end{equation}
in which we used the values $H_{d,th}/H=0.4$, $0.5$, $0.6$,
$0.7$, $0.8$ and $0.9$. None of the dotted lines is an acceptable fit
to the data. Close to the disc midplane, by eye inspection seems to
suggest that the dust disc scale-height
is $\sim 0.9H$ while one would estimate it to be of order $0.6H$
at $Z \sim 3H$. Both values are different from the dust disc 
scale-height returned by the least squares fit to the data,
$H_{d,th}^{fit}/H=0.84$. The physical reason for this behaviour is
clear: close to the midplane, where the gas density is high, the gas-dust
coupling is strong and the dust traces closely the Gaussian 
profile of the gas, whereas in the disc upper layers the coupling is
weak and the dust-gas ratio decreases as one moves further away
from the midplane.
These results 
emphasize the point that estimates of the dust disc scale-height
obtained using a Gaussian fit may lead to incorrect conclusions. 

Nevertheless, in order to compare with results published previously
in the literature, we measured the dust disc scale-height
in our simulations using such a fit. We found
$H_{d,th}/H=1.0$, $0.84$ 
and $0.40$, respectively, for $(\Omega\tau)_0=0.0001$, $0.001$ and
$0.01$. The variation of $H_{d,th}/H$ as a function of $(\Omega\tau)_0$
is shown by the diamonds in Fig.~\ref{hd_fit}. The dotted line
displays the function
\begin{equation}
H_d/H=0.7 \left( \frac{(\Omega\tau)_0}{0.001} \right)^{-0.2} \, ,
\end{equation}
which is the best fit to the data. Therefore, if we were to analyse
our simulations assuming that the dust density is Gaussian, we would
obtain $H_d \propto a^{-0.2}$ since $(\Omega\tau)_0 \propto
a$. Interestingly, this is not too different from the results of 
\citet{pinteetal08} who report an exponent equal to $-0.05$, although it
is clear that our results show a stronger relationship between the
dust disc scale-height and particle size. Both values are,
  however, largely different from the value $-0.5$ which is often
  reported in the literature
  \citep{dubrulleetal95,carballidoetal06}. This is because the latter
  is obtained when solving Eq.~(\ref{diff_advec_eq}) in the {\it
    strong settling limit $H_d \ll H$} that is mostly relevant for
  large particles. In this case, the vertical variation of $\Omega
  \tau_s$ can be neglected, the vertical profile is Gaussian and the
  exponent $-0.5$ is recovered. For the small particles we study here, the
  dust disc is thick and these vertical variations have to be taken
  into account, which leads to the more complicated expression given by
  Eq.~(\ref{dust_prof_eq}) and departure from a Gaussian profile.

\subsection{The Schmidt number}

The second result that emerges from our work is that a model having a
constant diffusion coefficient increasingly underestimates the dust
density as the particle size $a$ is decreased. In other words, the
vertically averaged dust diffusion coefficient decreases with
$a$. This is not unexpected, since the Schmidt number introduced in
Sect.~\ref{const_D} is known to be an increasing function of
$\Omega\tau$
\citep{cuzzietal93,schapler&henning04,youdin&lithwick07}. It is
unclear, however, whether such studies,
which assume homogeneous Kolmogorov--like turbulence,
are applicable to the highly magnetised flow of the corona. This is why we
did not attempt to make a direct comparison between our results and
these theories. For the sake of completeness, it is nevertheless 
instructive to report here the vertically averaged Schmidt number we
measured in each case. As described above, $\textrm{Sc}=1.5$ already
provides a good fit of 
the dust density in the case $(\Omega\tau)_0=10^{-2}$. For the cases
$(\Omega\tau)_0=10^{-3}$ and $10^{-4}$, we found that the vertically
averaged Schmidt numbers that best fit the data are respectively
$\textrm{Sc}=0.4$ and $\textrm{Sc}=0.03$\footnote{The fact that we
  obtain Schmidt numbers lower than one, in contrast to
  \citet{cuzzietal93}, \citet{schapler&henning04} and
  \citet{youdin&lithwick07} is due to
  our definition being different to that used in these theoretical
  studies, as discussed in detail by
  \citet{youdin&lithwick07}}. Although they indicate a scaling with 
$(\Omega\tau)_0$ close to linear, these results are not necessary in
disagreement with the result of \citet{youdin&lithwick07}, who found a
quadratic scaling, because of the vertical average we made when doing
such measurements. Note also that these fairly low values of the
  Schmidt number indicate that turbulent diffusion of the smallest
  dust particles can be much more efficient than angular momentum
  transport. This difference originates from their different physical
  origin: angular momentum is transported
  radially by the Maxwell and Reynolds stresses while
  dust diffuses away from the disc midplane because of the vertical velocity
  fluctuations of the gas. The former quantities decrease as one 
  moves away from the disc midplane,
  \citep[see][]{miller&stone00,fromang&nelson06} while the latter
  quatity increases away from the disc midplane. This means that angular
  momentum is inefficiently transported in the disc corona while small
  dust particles are efficiently diffused at the same location. As a
  consequence, the Schmidt number (i.e. the ratio of both diffusion
  coefficients) is much smaller than unity
  for the smallest of our particles.

\subsection{A toy model}

\begin{figure}
\begin{center}
\includegraphics[scale=0.5]{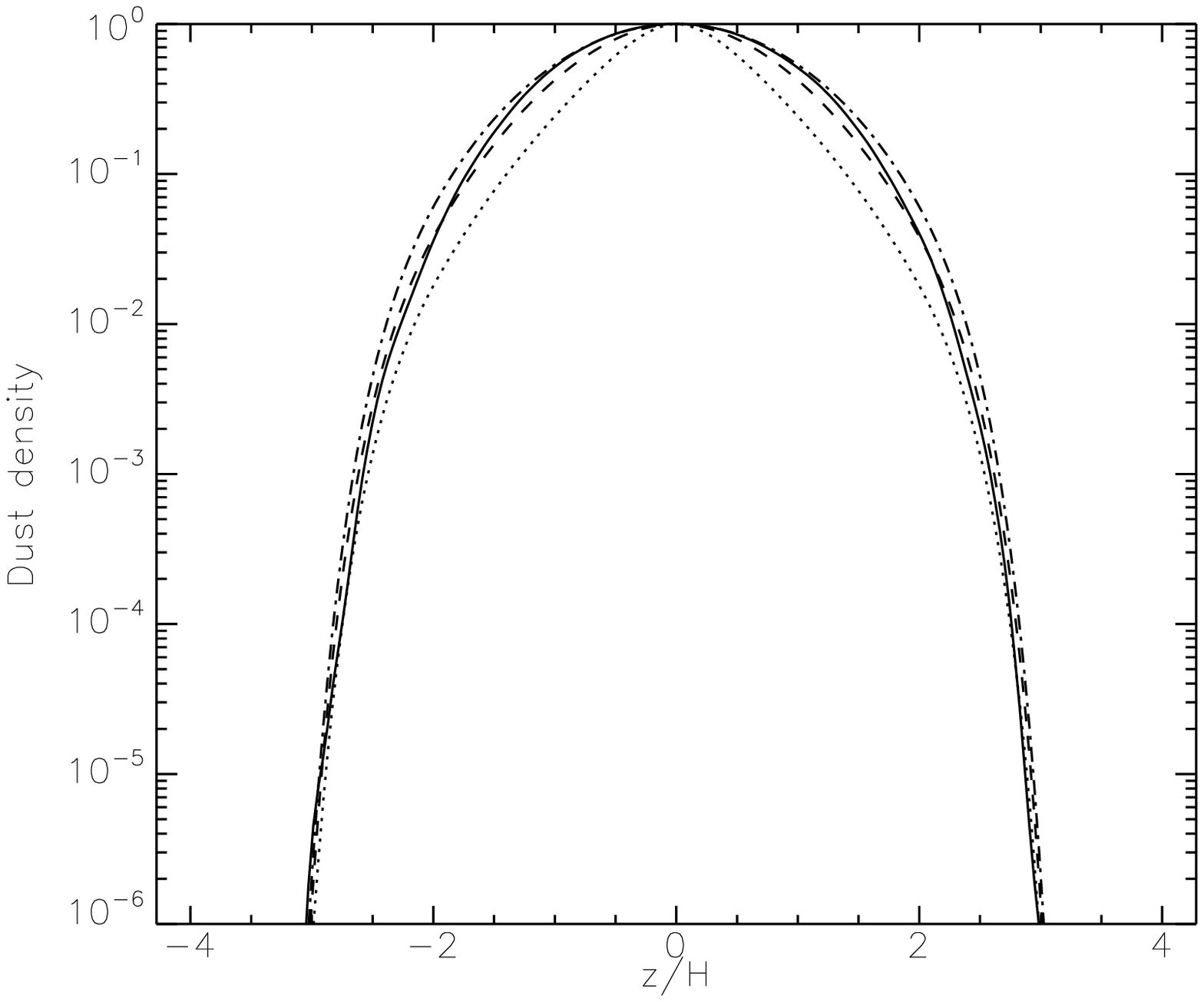}
\includegraphics[scale=0.5]{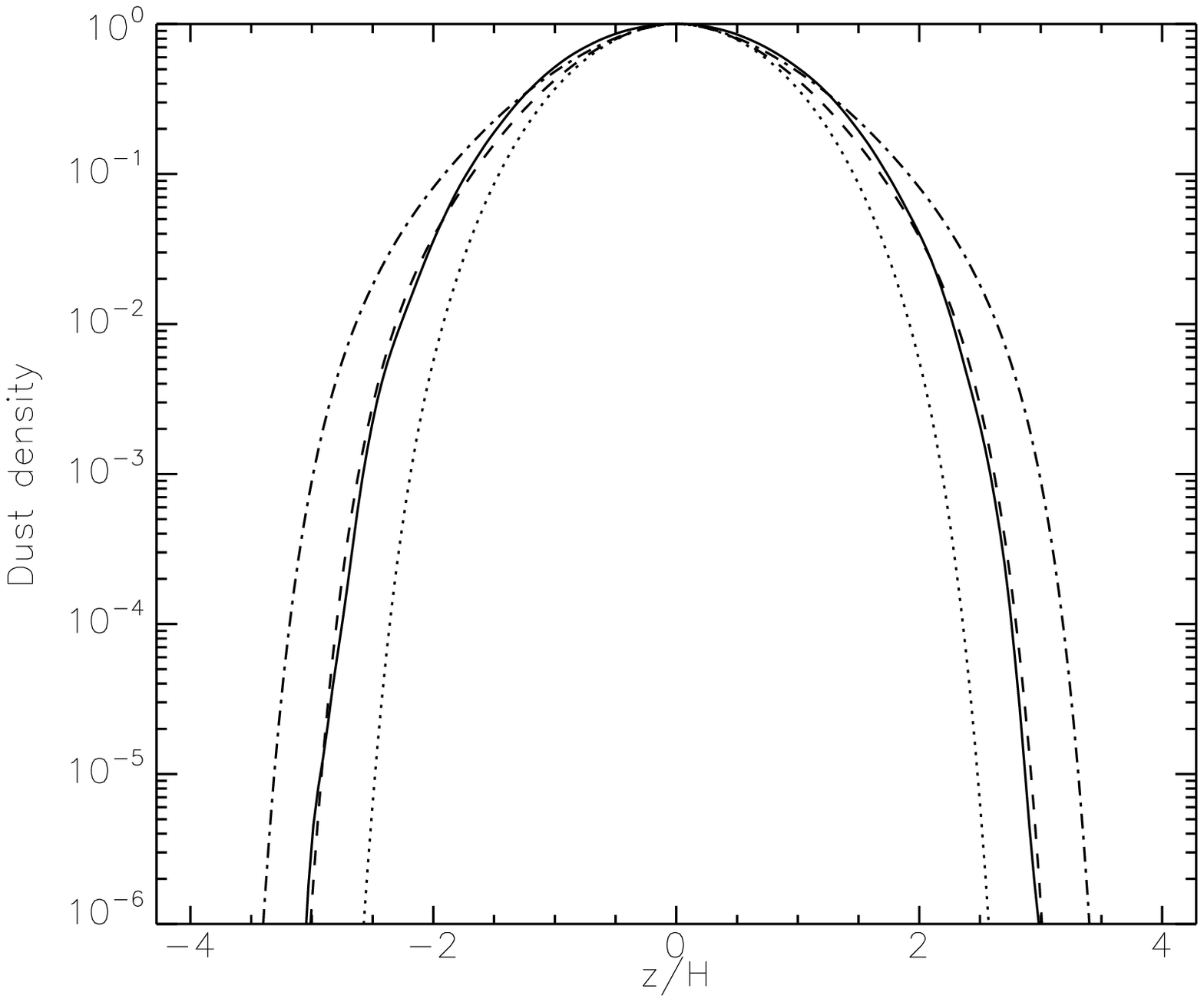}
\caption{Comparison between the dust density vertical profile (shown
  in both panels with the solid line) in the case
  $(\Omega\tau)_0=0.001$ with a simple toy model in which the
  turbulent velocity distribution is a step function (see text for
  details). In the upper panel, the velocity fluctuations are taken to
  be $\delta v_{z,up}/c_s \sim 0.15$ for $|z|>2H$, while $\delta
  v_{z,mid}/c_s=0.025$ ({\it dotted line}), $0.05$ ({\it dashed line})
  and $0.075$ ({\it dotted--dashed line}) for $|z|<2H$. In the lower panel, the
  velocity fluctuations are $\delta v_{z,mid}/c_s=0.05$ for $|z|<2H$
  and $\delta v_{z,up}/c_s \sim 0.075$ ({\it dotted line}), $0.15$
  ({\it dashed line}) and $0.30$ ({\it dotted--dashed line}) for
  $|z|>2H$. The results show that the dust density vertical profile is
fairly insensitive to the midplane velocity fluctuations but 
more strongly depends on their amplitude in the disc upper layers.}
\label{toy_fit}
\end{center}
\end{figure}

The main result of this paper is the construction of a simple model
that gives  
a reasonable fit to the simulations. In this model, the dust particles
are diffused away by turbulence with a diffusion coefficient that
scales with the square of the turbulent velocity
fluctuations. Accordingly, it should be possible in principle to
extract the vertical profile 
of the velocity fluctuations from the dust density vertical
profile. This is, however, an inversion problem. As such, it is
susceptible to 
being degenerate and we shall see in this section that this is indeed
the case.

The relevant question to ask in this context is the following:
provided we are able to measure the dust density vertical profile,
what is it mostly sensitive to? Can we hope to constrain the velocity
fluctuations in the disc midplane or is it mostly a consequence of
the amplitude of the fluctuations in the upper layers. To answer that
question, we designed the following toy model: guided by the 
vertical profile of the velocity fluctuation amplitudes
shown in Fig.~\ref{vel_fluc}, we
considered the following analytic vertical profile for the 
velocity fluctuations:
\begin{equation}
\delta v_z= \left\{ \begin{array}{ll} 
   \delta v_{z,mid}+[\delta v_{z,up}-\delta v_{z,mid}]\left(\frac{|z|}{2H}\right)^2 \textrm{ if } |z|< \textrm{2H} \\
   \delta v_{z,up} \textrm{ otherwise}
   \end{array} \right.
\label{dvz toy profile}
\end{equation}
where $\delta v_{z,mid}$ and $\delta v_{z,up}$ stand for the turbulent
velocity fluctuations in the disc midplane and in the disc corona. As
shown on Fig.~\ref{vel_fluc}, typical numerical values are $\delta
v_{z,mid}/c_s \sim 0.05$ and $\delta v_{z,up}
/c_s \sim 0.15$. To investigate the sensitivity of the results to the
midplane velocity fluctuations, we calculated the dust
density vertical profile in the case $(\Omega\tau)_0=0.001$ by
numerically integrating Eq.~(\ref{diff_advec_eq_II}), using Eq.~(\ref{D_vel})
 and (\ref{dvz toy profile}) with $\delta
v_{z,mid}/c_s=0.025$, $0.05$ and $0.075$ and $\delta
v_{z,up}/c_s=0.15$ (i.e. we used in the disc corona the value
suggested by the simulation data) . The results are summarized on the
upper panel of Fig.~\ref{toy_fit}, where the dust density profile
in the case $(\Omega\tau)_0=0.001$ is shown with the solid line, while the
numerically integrated profiles are represented by
dotted, dashed and dot--dashed lines for $\delta
v_{z,mid}/c_s=0.025$, $0.05$ and $0.1$, respectively. 
The last three curves are
very similar in this plot and all give a fairly good fit to the
simulations, especially in the disc upper layers. This shows that the
dust density vertical profile is
fairly insensitive to the midplane velocity fluctuations. To estimate
the sensitivity of the dust density profile to the velocity
fluctuations in the upper layers, we repeated the same analysis using $\delta
v_{z,mid}/c_s=0.05$ (i.e. we used in the disc midplane the value
suggested by the simulation results) and $\delta v_{z,up}/c_s=0.075$, $0.15$ and
$0.30$. The results are shown on the bottom panel of 
Fig.~\ref{toy_fit} with the same conventions as for the upper panel. In
this case, the dust density vertical profile is seen to be much more
sensitive to the upper layer velocity fluctuations and only the dashed
curve, for which $\delta v_{z,up}/c_s=0.15$ (i.e. in rough agreement with
the numerical data), is in good agreement with the data.

The implications of these results are twofold. First, the
simple toy model described by Eq.~(\ref{dvz toy profile}) would be
suitable to use when trying to fit the observations as it reproduces the
numerical data fairly well if we choose the values $\delta
v_{z,mid}/c_s=0.05$ and $\delta v_{z,up}/c_s=0.15$, compatible with
the simulations. But these results
also show that such a fit would only provide sensitive information
about the turbulent velocity fluctuations in the disc upper layers.
The physical reason for this
last point is that such a fit is mostly sensitive to the
properties of the region where the gas and dust decouple. For the small
particles studied in this paper (and observed using the Spitzer
telescope), this region turns out to lie 
in the disc upper layers. When observing larger particles at longer
wavelengths (for example with ALMA), it will become possible to constrain
the turbulent velocity fluctuations of the disc closer to the 
midplane as such particles will settle and decouple deeper in the disc.

\subsection{Limitations and future prospects}
\label{limitations}

Of course, there are strong limitations to our work due to the
complex and CPU-intensive nature of global simulations 
of protoplanetary discs. On the
purely numerical side, the limited resolution we used is of course an
issue, as was pointed recently in a number of studies
\citep{fromang&pap07, simonetal08}. Proper simulations should
include explicitly microscopic diffusion coefficients (viscosity and
resistivity), as the latter have been shown to be important in
determining the saturation level of the turbulence
\citep{fromangetal07,lesur&longaretti07}. However, the resolutions 
required to include these processes in global simulations are currently out of
reach and one must instead 
rely on the subgrid model provided by numerical dissipation to carry
out global numerical simulations. On a more physical side, there are
also limitations due to the simple disc model we used. The locally
isothermal equation of state we used is not appropriate for the disc inner
parts that we are simulating as the gas there is optically thick. 
This could have numerous effects. For example, \citet{dullemond02} and
\citet{dalessioetal98} report a temperature increase in the inner disc
upper layers. Such an increase would strengthen the coupling between
  gas and grains in the disc corona (through  
  a decrease in the local value of the parameter $\Omega
  \tau_s$). This would cause small particles to settle less than
  reported in this paper and could change the relationship between $H_d$ 
  and $a$. Obviously, the significance of the comparison we tentatively
  made between our simulations and the observations of
  \citet{pinteetal08} should be taken with care.
Another topic of concern in our simulations is the
  assumption of ideal MHD. It is indeed well known that protoplanetary
  discs are so poorly 
  ionized because of their large densities and low temperatures that parts
  of the flow, refered to as dead zones, remain laminar
  \citep{gammie96}. We completely ignored the effects of dead zones in
  the present paper. Clearly, future work should improve
the thermodynamic treatment of the gas, possibly including radiative
transfer and dead zones.

Nevertheless, we have shown in this paper that dust observations can be
used in principle to constrain the properties of MHD turbulence in
discs. We have found that even the simplest simulations provide
disagreements with previously used fits and diffusion models because
of the nature of disc turbulence. This illustrates even further the need to
compare directly observations and numerical simulations. It will be
important in the future to generate a grid of more realistic discs
models (varying the disc parameters, including dead zones, flaring
discs, non isothermal discs) and produce synthetic observations that
could be compared in the next few years with multiwavelengths
observations. This comparison would provide diagnostics of disc
turbulence, the existence (or not) of dead zones and thus constrain
planet formation models. The recent work of \citet{pinteetal08} shows
that such multiwavelengths observations are starting to become
feasible. When combined with future instruments like Herschel and
ALMA, large samples will become available and will provide a wealth of
constraints on disc structure and properties when combined with
appropriate global numerical simulations of protoplanetary discs.

\section*{ACKNOWLEDGMENTS}
The simulations presented in this paper were
performed using the QMUL High Performance Computing Facility purchased
under the SRIF initiative. It is also a pleasure to acknowledge fruitful
discussions with J--C.Augereau and P--Y. Longaretti. We also thank an
anonymous referee whose comments significantly improved the paper. 

\bibliographystyle{aa}
\bibliography{author}

\end{document}